\pdfoutput=1
\documentclass[11pt]{article}

\usepackage[]{acl}
\usepackage{times}
\usepackage{latexsym}

\usepackage[T1]{fontenc}
\usepackage[utf8]{inputenc}

\usepackage{microtype}

\usepackage{inconsolata}

\usepackage{subcaption}
\usepackage{gensymb}
\usepackage{titlesec}
\usepackage[english]{babel}
\usepackage{booktabs}
\usepackage{subfiles}
\usepackage{amsmath}
\usepackage{tabularx}
\usepackage{comment}
\usepackage{siunitx}
\usepackage{pdflscape}
\usepackage{multirow}
\usepackage{amssymb}
\usepackage[most]{tcolorbox}

\usepackage{tabularx}
\usepackage{graphicx}
\usepackage{pifont}
\usepackage{latexsym}
\usepackage{datetime}
\usepackage{hyperref}
\usepackage{adjustbox}
\usepackage{color, soul}
\usepackage{enumitem}

\newdate{date}{01}{05}{2022}
\newdate{date1}{15}{02}{2022}
\usepackage{url}
\usepackage{comment}
\usepackage{caption}
\usepackage{underscore}
\usepackage{diagbox}
\usepackage{tabularx}
\renewcommand{\vec}[1]{\mathbf{#1}}

\usepackage[show]{chato-notes}
\usepackage{multirow}
\usepackage[skip=0.333\baselineskip]{caption}
\usepackage{subcaption}
\usepackage{enumitem} 
\newcommand{\para}[1]{\paragraph{\textbf{#1}.}}

\newcommand{\iclnz}{0-shot}

\newcommand{\uls}{\begin{itemize}[leftmargin=*]}
\newcommand{\ule}{\end{itemize}}
\newcommand{\ols}{\begin{enumerate}[leftmargin=*]}
\newcommand{\ole}{\end{enumerate}}

\usepackage{amssymb}
\newcommand{\cmark}{\ding{51}}%
\newcommand{\xmark}{\ding{55}}%
\usepackage[utf8]{inputenc}
\usepackage{url}
\usepackage[table]{xcolor}
\usepackage{colortbl}
\definecolor{lightblue}{RGB}{173, 216, 230}

\definecolor{mygray}{gray}{0.9} % 0 = black, 1 = white. So 0.9 is light gray

\newcommand{\gc}[1]{\cellcolor{mygray}#1}

\newcommand\mybox[2][]{\tikz[overlay]\node[fill=blue!20,inner sep=2pt, anchor=text, rectangle, rounded corners=1mm,#1] {#2};\phantom{#2}}

\newcommand{\bbox}[1]{\mybox[fill=blue!20]{#1}}
\newcommand{\rbox}[1]{\mybox[fill=red!20]{#1}}

\title{Mask-to-Correct$^+$: Leveraging Retriever Diversity for Masking-guided Faithful Fact Correction}

\author{Payel Santra\thanks{These authors contributed equally to this work.}, Lavisha Sharma\footnotemark[1], Madhusudan Ghosh \and Partha Basuchowdhuri \\
Indian Association for the Cultivation of Science, Jadavpur, Kolkata-700032, India \\ \{\href{mailto:payel.iacs@gmail.com}{payel.iacs}, \href{mailto:lavishasharma2612@gmail.com}{lavishasharma2612}, \href{mailto:madhusuda.iacs@gmail.com}{madhusuda.iacs}\}@gmail.com \\ \and \href{mailto:partha.basuchowdhuri@iacs.res.in}{partha.basuchowdhuri@iacs.res.in}}
\begin{document}
\maketitle
\begin{abstract}
{
The rapid spread of misinformation on social media highlights the need for robust, automated fact correction frameworks. However, existing works rely on supervised learning from manually annotated claim-evidence pairs, which are scarce and prone to biases, limiting their generalization across domains. 
Moreover, these methods overlook semantic faithfulness in their correction process. To address these challenges, we propose \textbf{Mask-to-Correct (M$_2$C)}, a training-free, inference-only Retrieval Augmented Generation (RAG) based framework that leverages \emph{diversity-aware masking} to identify erroneous spans of claims and evaluate the faithfulness of corrections using retrieved evidence. However, the effectiveness of RAG heavily depends on the choice of retriever, which may vary across queries. To mitigate this, we further introduce \textbf{M$_2$C$^+$}, an ensemble-based framework that combines corrections across multiple rankers to reduce retrieval bias and improve robustness. Extensive experiments on the benchmark datasets demonstrate that our proposed frameworks consistently outperform all baselines, achieving up to 14\% improvement in SARI scores, without using gold evidence.
% Our proposed model \textbf{M$_2$C$^+$} outperforms all baselines in fact correction without using gold evidence.
%
% The rapid spread of misinformation on social media highlights the need for automated fact correction systems. Existing works rely on supervised learning from manually annotated claim–evidence pairs, which are scarce and prone to bias, limiting their generalization across domains. Moreover, these methods often overlook semantic faithfulness in their corrections. To address these challenges, we propose Mask-to-Correct (M$_2$C), a training-free Retrieval-Augmented Generation (RAG) framework that leverages diversity-aware masking to identify erroneous spans within claims and generate faithful corrections guided by retrieved evidence. However, the effectiveness of RAG depends heavily on retriever choice, which can vary across queries. 
% To mitigate this, we further introduce M$_2$C$^+$, an ensemble-based extension that aggregates corrections across multiple retrievers to reduce retrieval bias and enhance robustness. Extensive experiments on the FEVER and SciFact datasets show that M$_2$C$^+$ consistently outperforms all supervised and unsupervised baselines, achieving up to 7\% improvement in SARI and higher BART scores, demonstrating superior factuality and faithfulness without requiring gold evidence.
}
\end{abstract}
%
%
%
% . Such datasets are limited in number, resource-intensive to build, and prone to biases, limiting their generalization across domains. 
%  However, existing approaches largely depend on supervised learning with manually annotated claim-evidence pairs, which are scarce, costly to create, and often biased—limiting their applicability across domains.
 %% Experimental results on benchmark datasets demonstrate that our proposed approach, \textbf{M$_2$C$^+$}, consistently outperforms existing baselines for the fact correction task without relying on ground-truth evidence.
 %
% , we propose \textbf{Mask-to-Correct (M$_2$C)}, a training-free, Retrieval-Augmented Generation (RAG)-based framework for faithful factual error correction. Our method incorporates a diversity-aware masking strategy to identify erroneous spans, and assesses the faithfulness of each correction based on its consistency with the evidence.
%

\section{Introduction}

% Large language models (LLMs) have made a significant impact in natural language processing (NLP), achieving state-of-the-art performance across a wide range of tasks, including text generation~\cite{sun2023text}, summarization \cite{fu2023gptscore,liu2022brio}, and fact verification \cite{yue2024retrieval}. 

Large language models (LLMs) have made a significant impact across a wide range of tasks~\cite{sun2023text,fu2023gptscore,yue2024retrieval}.
However, LLMs \cite{touvron2023llama,wang2022gpt} have intensified the generation of misinformation with the help of suitably crafted adversarial prompts \cite{zou2023adverserial}.
The topically coherent and fluent nature of LLM-generated text \cite{liu2021gpt} potentially makes it even harder to detect any injected misinformation \cite{positionalbias}. On the other hand, the inherent susceptibility of language models towards hallucinations~\cite{ji2023towards} often results in factually inconsistent outputs.

% To address these challenges, we need to develop systems that can assist in detecting misinformation and generate a corrected version of the sentence.

% Large language models (LLMs) have made a significant impact across various NLP tasks~\cite{sun2023text,fu2023gptscore,liu2022brio,yue2024retrieval}, but they also raise concerns by enabling the generation of topically coherent but potentially misleading content~\cite{liu2021gpt,zou2023adverserial}. Adversarial prompts and the inherent tendency of LLMs make it even harder to detect any injected misinformation~\cite{positionalbias,ji2023towards}.  To address these challenges, we need to develop systems that can assist in detecting misinformation and generate a corrected version of the sentence.
\begin{figure}[t]
    \centering
\includegraphics[width=0.85\columnwidth]{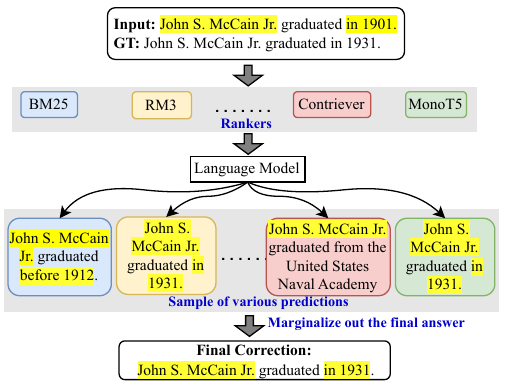}
    \caption{\small An illustrating example of our proposed model. Given an incorrect claim, each retriever yields a different set of evidence, leading to diverse corrections. Our approach encompasses these corrections to marginalize out a factually correct version of the claim. Here, `GT’ denotes the ground-truth claim.}
    \vspace{-1em}
    \label{fig:motiv_pic}
\end{figure} 

% automated systems that not only detect but also correct such errors.
In this paper, we primarily address the task of factual error correction, which involves editing a given input sentence (or claim) to make it factually consistent with the supporting evidence. Unlike the fact verification task,  which involves classifying a claim's veracity class (``TRUE'', ``FALSE'', etc.)~\cite{schuster2019towards,schuster2021get,jiang2021exploring} based on supporting or refuting evidence~\citep{asai2023retrieval}, fact correction extends this task by revising incorrect claims into correct ones while preserving their semantics. Most standard approaches of this task focus on preserving grammar and fluency using fully or distantly supervised approaches~\cite{shah2020automatic,thorneEvi,he2024improving}, but ignore faithfulness\footnote{Semantic faithfulness refers to preserving a claim’s core meaning, its proposition, structure, and intent while modifying only factually incorrect spans based on external evidence. Unlike grammaticality or fluency, which focus on surface form, semantic faithfulness ensures the integrity of meaning with respect to both the original claim and the supporting evidence.} to the supporting evidence. Moreover, these methods require annotated evidence of each claim, which is resource-intensive and time-consuming and also susceptible to pooling and exposure biases~\citep{yang2019unsupervised} (see Appendix \ref{appx:annotation}).

% This process enumerates a set of candidate corrections for the claim and selects the best corrected claim.

% estimates the veracity of claims in form of a classification task, \cite{schuster2019towards,schuster2021get,jiang2021exploring} by using evidence for or against them~\citep{asai2023retrieval}, the factual error correction task revises incorrect claims into correct ones while preserving their semantics and consistency. 
% This makes this task practically more suitable to real-world requirements.
% Fact verification works on predicting the veracity class (``TRUE'', ``FALSE'', etc.) of a claim. Whereas, fact correction extends this task to enumerate candidate corrections for the claim and selects the best corrected claim.

% A common baseline is to directly prompt an LLM to ``fix'' the claim.

Inspired by~\citet{huang2023zero}, we address these challenges by introducing training-free approaches, such as zero-shot prompting followed by a Retrieval-Augmented Generation (RAG)-based framework~\cite{lewis2020retrieval,izacard-grave-2021-leveraging}, which enhances an LLM's response generation by integrating external knowledge from large-scale corpora, thereby eliminating the need for manually annotated ground-truth evidence. However, such prompting approaches face some challenges:
%
% \begin{itemize}
%     \item \textbf{Ensuring minimal change:}The corrected claim should preserve the original semantics. Identifying potentially false or weakly supported spans within a claim may help us address it convincingly.
%     \item \textbf{Retriever quality dependency:} As illustrated in Figure~\ref{fig:motiv_pic}, the effectiveness of RAG-based correction depends heavily on the quality of the retrieved evidence. A single retriever may fail to capture diverse or ambiguous contexts and accumulating complementary evidence from multiple retrievers leads to more reliable corrections.
% \end{itemize}
\textbf{Ensuring minimal change:} The corrected claim should preserve the original semantics. Identifying potentially false or weakly supported spans within a claim may help us address it convincingly.
\textbf{Retriever quality dependency:} As illustrated in Figure~\ref{fig:motiv_pic}, the effectiveness of RAG-based correction depends heavily on the quality of the retrieved evidence. A single retriever may fail to capture diverse or ambiguous contexts and accumulating complementary evidence from multiple retrievers leads to more reliable corrections.

To overcome these challenges, we propose a novel masking aware, three-stage framework M$_2$C (\underline{M}ask-\underline{to}-\underline{C}orrect) that identifies and masks erroneous spans in a claim, generates corrected candidates using retrieved evidence, and selects the most faithful revision through a semantic–factual scoring function. We further introduce M$_2$C$^+$, an ensemble-based extension that aggregates multiple retrieval generation pathways for improved robustness.
% The novelty of this approach lies in a task-specific integration that enables training-free and evidence-agnostic fact correction. 
% Unlike QA or summarization, fact correction requires localization of factual errors while preserving correct spans. 
Importantly, to our knowledge, our work is the first to integrate RAG with a diversity-aware masking strategy that ensures broader coverage of plausible error regions while minimizing unnecessary edits, hallucinations, thereby increasing faithfulness. Thus, the novelty of this approach lies in a task-specific, lightweight aggregation across different retrievers that enhances correction stability without additional training overhead. Furthermore, to strengthen our study, we conducted extensive analyses in multiple dimensions, including the impact of different masking strategies, retriever combinations, computational efficiency, and correction scoring schemes, along with detailed sensitivity and ablation studies. 
% ///////
% To obtain diverse examples,
% we propose to rerank a top-retrieved set of candidate examples
% with maximal marginal relevance (MMR) [3], which is a greedy
% algorithm that seeks to maximize the similarity of the selected
% examples with the input, while simultaneously minimizing the
% similarity between the selected examples
%/////////
% Moreover, we incorporate a faithfulness-preserving control mechanism, which ensures minimal semantic drift and reduces hallucination.

% precise localization of factual errors while preserving grammaticality and semantic intent.
% Our diversity-aware masking constrains LLM edits to plausible error spans, retrieval grounding ensures factual alignment without gold evidence, and aggregation stabilizes predictions across retrieval variations. Together, these yield a unified framework that systematically controls LLM behavior for faithful and minimally invasive correction.

Our contributions are summarized as follows: 1)~We introduce a training-free fact correction framework,  M$_2$C, that does not require gold evidence. 2) We propose a diversity-aware masking strategy to identify high-impact erroneous spans more effectively than traditional methods. 3) To the best of our knowledge, we are the first to introduce majority voting-based ensembling approach for the downstream task through our framework  M$_2$C$^+$.

\section{Related Work}

\para{Fact Verification}
Early works on fact verification explored supervised methods using pre-trained models~\citep{stammbach-neumann-2019-team,krishna2022proofver,soleimani2020bert, chernyavskiy-ilvovsky-2019-extract}, multitask learning~\cite{hidey2018team,lewis2020retrieval}, and Graph-based learning~\cite{zhou-etal-2019-gear, liu-etal-2020-fine}. Recent methods have deployed LLMs with in-context learning (ICL) and RAG to improve scalability and factual grounding~\cite{Kojima2022LargeLM,santra2024absence,min-etal-2022-metaicl,santra2025curious}.

% To enhance performance further, claim decomposition, iterative retrieval, and evidence re-ranking have been explored, addressing limitations like insufficient context and hallucination.

 % Transformer-based models, including BERT, were explored for both retrieval and verification, incorporating pointwise and pairwise loss functions with hard negative mining~\cite{soleimani2020bert, chernyavskiy-ilvovsky-2019-extract, nie-etal-2019-revealing, portelli-etal-2020-distilling}. Hybrid approaches combined TF-IDF and named entity-based search with transformer-based models~\cite{malon2019team}. Graph-based methods, such as GEAR and KGAT, introduced structured reasoning by modeling evidence relationships via graph attention networks~\cite{zhou-etal-2019-gear, liu-etal-2020-fine}. There is also a focus on using the web for evidence retrieval and unsupervised methods to reduce annotation costs~\citep{subramanian-lee-2020-hierarchical,stammbach-2021-evidence}.

\para{Retrieval Augmented Generation}
% RAG combines retrieval with generation to incorporate relevant, up-to-date information during decoding, improving factual accuracy without task-specific fine-tuning. Advances include better retrieval via re-ranking \cite{glass2022re2g}, adaptive context selection, and iterative refinement, as well as improved integration through relevance-based weighting.
RAG-based methods integrate retrieval from external KBs to reduce hallucinations during generation~\cite{guu2020retrieval,shi2023replug,Lan2023CopyIA,jiang2023active,zhang2023repocoder,izacard2020leveraging,borgeaud2022improving}. Recent advances focus on improving retrieval through adaptive context selection~\cite{jeong-etal-2024-adaptive}, iterative refinement, and relevance-based integration~\cite{glass2022re2g}. A similar work, EoR~\cite{li2024unraveling}, adaptively fuses retrievals from different sources to reduce evidence-level inconsistencies in retrieval-augmented QA task. In contrast, our framework performs correction-level ensembling across independently corrected outputs using various retrievers.

\para{Fact Correction}  
Existing supervised strategies include methods that target factual inconsistencies in text summarization~\cite{fabbri2022improving,adams2022learning}, employ masking techniques to guide correction~\cite{shah2020automatic,thorne2021evidence}, or perform iterative edits based on predicted truthfulness scores~\cite{chen2023converge}. In some studies, distantly supervised correctors were trained on synthetic data, avoiding the need for explicit masking~\cite{he2024improving,ashok2023scifix}. Recent studies have addressed this problem with unsupervised approaches by leveraging a QA-based pipeline~\cite{huang2023zero} with knowledge graph-based retrieval~\cite{bayat2023fleek}.

% Unsupervised approaches like ZEROFEC and FLEEK use QA-based pipelines, leveraging task-specific models, domain adaptation, and knowledge graphs to ensure faithful fact correction and verification.

% Prior work on fact correction has explored both supervised and unsupervised strategies. C{\tiny OMP}E{\tiny DIT}~\cite{fabbri2022improving} and R{\tiny EVISE}R{\tiny EF}~\cite{adams2022learning} focus on eliminating factual errors in summaries. MaskCorr~\cite{shah2020automatic} and T5-Distant~\cite{thorne2021evidence} introduce masking strategies paired with a fine-tuned T5 corrector for context-aware infilling. VENCE~\cite{chen2023converge} applies iterative edits guided by predicted truthfulness scores. LIFE~\cite{he2024improving} proposes a distantly supervised corrector trained on synthetically corrupted data without relying on explicit masking. SCIFIX~\cite{ashok2023scifix} addresses scientific claims using synthetic claim-aware decoding. Among unsupervised approaches, Z{\tiny ERO}F{\tiny EC}~\cite{huang2023zero} uses a zero-shot QA-based framework with task-specific LMs and domain adaptation to ensure faithfulness. FLEEK~\cite{bayat2023fleek} leverages knowledge graphs and web evidence in a similar QA-driven pipeline for both correction and verification.
% \begin{figure*}[t]
%     \centering
% \includegraphics[width=\columnwidth]{figure/archt.drawio_cropped.pdf}
%     \caption{An illustration 
%     }
%     \label{fig:icl_prp}
% \end{figure*}

\begin{figure*}[t]
    \centering
    \includegraphics[width=0.99\textwidth]{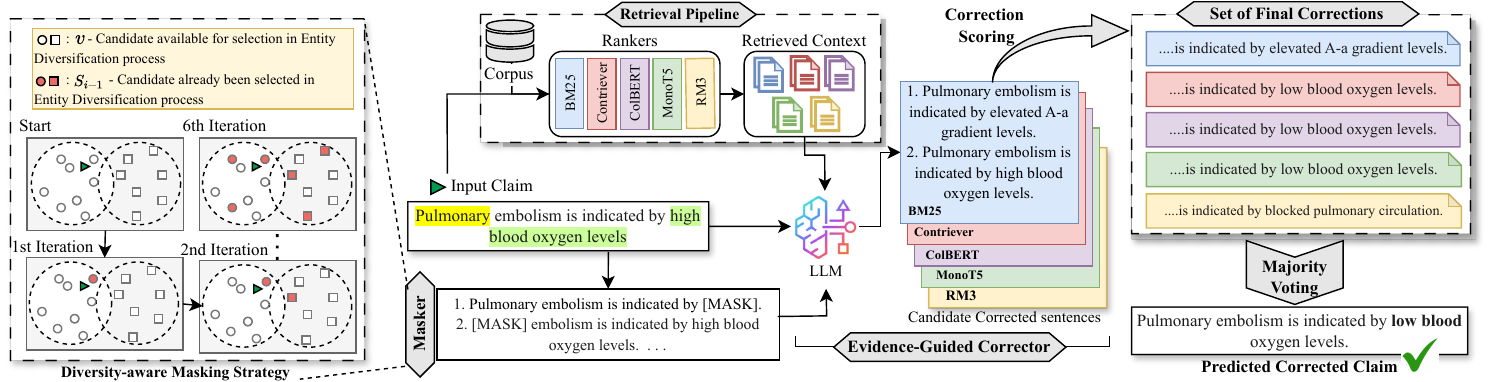}
    % {figure/Fact_correction-archit.pdf}%{figure/ARchit.drawio_cropped.pdf}
    \caption{\small Schematic overview of our proposed M$_2$C$^+$. Given a claim, the diversity-aware masker (an iterative module) identifies and prioritizes spans that are likely to be incorrect. It selects entities based on similarity and diversity (e.g., selecting the most similar entity in the 1$^{\text{st}}$ iteration (from the same cluster); in the 2$^{\text{nd}}$ iteration, it adds an entity that remains relevant to the claim but is diverse from the one previously selected (e.g., from a different semantic cluster); and so on.). Here, we have highlighted top-2 entities for masking. Top documents are then retrieved from an external corpus using multiple retrievers. Each retrieval path is used by an LLM to generate corrected candidates. A correction scoring module selects the best correction per retriever, and majority voting is applied across these outputs to produce the final corrected claim.}
\vspace{-1em}
    \label{fig:archit}
\end{figure*}

In contrast to previous studies, we adopt a masking-guided unsupervised approach for the fact correction task by integrating an ensemble-based RAG framework to ensure both faithfulness and factuality. Unlike traditional maskers, we proposed a diversity-aware masking strategy.

% \vspace{-5mm}

\section{Methodology}
% \input{fig_def/mmr_archit}
% \input{fig_def/fig2_archi}
% Adversarial fabrication and hallucination of information, both have become growing concerns with the rise of large language models (LLMs). 

% The task of factual error correction 

% In this paper, our task is faithful fact correction, involves revising an incorrect claim into a correct one by utilizing contexts for or against it. Given an input claim $\vec{x}$, our objective is to introduce an automated system $\phi$, to rectify the factual errors and produce a revised claim $\hat{\vec{x}}$ that is factually consistent with the provided evidence $\mathcal{E}(\vec{x})$. 
In this paper, we address the task of faithful fact correction, which involves rectifying an incorrect claim using contextual evidence. Formally, given an input claim $\vec{x}$ (correct, incorrect or ambiguous), our goal is to design an automated system $\phi$ that identifies and rectifies factual errors (if they exist) to generate a revised claim $\hat{\vec{x}}$ that is consistent with the associated evidence $\mathcal{E}(\vec{x})$.
More formally, $\phi: (\vec{x}, \mathcal{E}(\vec{x})) \mapsto \hat{\vec{x}}$. We hypothesize that the predicted claim $\hat{\vec{x}}$ should be grammatically correct and minimally altered to preserve the original semantics of $\vec{x}$. Furthermore, our method is designed to maintain factual consistency if the $\vec{x}$ is already correct. We propose a novel three-stage masking-guided, training-free, and annotation-free framework, M$_2$C, that leverages the reasoning capabilities of LLMs. The key components are described as follows:

% To achieve this, we propose Mask-to-Correct (M₂C) — a training-free and annotation-free framework that leverages the reasoning capabilities of large language models (LLMs).
% M₂C decomposes the correction process into three sequential stages:

% Masking: identify and mask potentially incorrect spans in the input claim;

% Correction: generate candidate corrections using retrieved evidence; and

% Scoring: select the most faithful and factually consistent correction.

% We further extend this to M₂C+, which aggregates candidate corrections from multiple retrievers to mitigate retrieval bias and enhance robustness (Figure 2).
% \section{Methodology}
% In recent times, rapid spread of misinformation over social media has become a practice for different interest groups. In such cases, trying to find evidence to corroborate the truthfulness of a claim is a challenging task. Furthermore, obtaining reliable ground-truth annotations for factual correction is labor-intensive and costly, and also likely to suffer from pooling and exposure biases~\citep{yang2019unsupervised} (see Appendix \ref{appx:annotation}).

% \vspace{2mm}
% Additionally, if a claim is factually correct, then our framework does not prefer any modification. Our proposed framework for factual error correction, \underline{M}ask-\underline{t}o-\underline{C}orrect (M$_2$C), consists of three key components: 1) Masker, 2) Evidence-Guided Corrector, and 3) Correction Scoring. Described in the following subsections. 

\subsection{Masker} \label{sec:masker}
The objective of this module is to identify salient and diverse spans within a claim that are most likely to require factual revision, while preserving the overall structure and semantics of the input without explicitly verifying correctness. Formally, given an input claim $\vec{x}=\{x_i\}_{i=1}^n$, where $x_i$ denotes the $i^{th}$ token, the module first extracts a set of candidate entities or spans $E_s(\vec{x})=\{e_i\}_{i=1}^s$, with $s<n$. Each entity $e_j$ is then independently masked to produce a set of perturbed claims $\{\tilde{\vec{x}}_j\}_{j=1}^s$, where $\tilde{\vec{x}}_j$ is obtained by replacing $e_j$ with a $[MASK]$ token. However, standard masking strategies such as random masking~\cite{thorneEvi} or purely entity based masking~\cite{huang2023zero} either lack in semantic focus or rely solely on syntactic boundaries, often missing contextually important spans and leading to suboptimal corrections. To address this limitation, we propose a Maximal Marginal Relevance (MMR)~\cite{carbonell1998use} based \textbf{diversity-aware masking paradigm} to iteratively re-rank candidate entities. This yields a prioritized subset $\mathcal{R}_m=\{r_i\}_{i=1}^m$, where $m \leq s$, ensuring that the selected spans collectively capture diverse and informative aspects of the claim for our downstream task.

As illustrated in Figure~\ref{fig:archit}, using MMR algorithm, we iteratively add those entities in an incrementally growing list that are both relevant to the claim and diverse with respect to one another. 
% Specifically, MMR is an iterative
% algorithm which maintains an incrementally growing list of selected
% instances and the score is a function of
%
% To understand the iterative process of MMR, observe the expanation of Masker in Figure \ref{fig:archit}, observe we first pick the most similar entity (from the same cluster as input claim), then add next, which is both relevant to the claim and diverse with respect to just recently selected (i.e., sleceted from different cluster than the claim and 1st entity), and so on, hence we get a incrementally growing list $\mathcal{R}_m$.
%
% As illustrated in Figure~\ref{fig:archit}, the Masker iteratively selects entities using the MMR criterion—starting with the entity most similar to the input claim, followed by others that remain relevant yet diverse with respect to previously chosen ones. 
%
 % Unlike entity masking, \textit{our masking strategy incorporates semantic relevance and diversity using the Maximal Marginal Relevance (MMR) approach}~\citep{carbonell1998use} to provide a prioritized order for masking. For a given claim $\vec{x}$, let $E_s(\vec{x})$ be the set of $s$ entities and phrases, where $s<n$. We apply the iterative process MMR to select and re-rank top-$m$ $(<s)$ entities, $\mathcal{R}_m$. 
 %
 % Thus, we get a incrementally growing list of selective instances by using MMR scores, which
 %
 More formally, it is a function of the following: (a) the iteration step $i<m$, ($m$ - the number of entities to be selected), (b) the set of entities selected until the $(i-1)^{th}$ step, $S_{i-1} \subseteq \mathcal{R}_m $, and (c) a hyperparameter $\alpha \in [0,1]$ that balances relevance with diversity. Formally, $\forall v \in \{E_s \setminus S_{i-1}\}$, 
 
{\small
\begin{equation}
\theta_{\text{MMR}}(\vec{x}, v, S_{i-1}; \alpha) = \alpha \, \theta(\vec{x}, v) 
- (1 - \alpha) \max_{s \in S_{i-1}} \theta(v, s),
\label{eq:mmr}
\end{equation}
}

% \endgroup
where, $\theta$ is the similarity function.
For the claim mentioned in Figure \ref{fig:archit}, the method first selects the most relevant one (e.g., ` high blood oxygen levels'), and then selects diverse but relevant entities (like `Pulmonary') by ignoring those similar to the already selected ones.

\subsection{Evidence-Guided Corrector}
% After generating masked sentences $\{\vec{\Tilde{x}}_j\}_{j=1}^k$ from a given input claim $\vec{x}$, in this module, we introduce a RAG-based pipeline to predict those masked sentences using a given set of evidence $\mathcal{E}(\vec{x})$. Here, the model directly leverages a small set of examples that are similar to the current claim from an external corpus $\mathcal{T}$, eventually including these with the masked claim as a part of an input prompt to an LLM~\cite{liu2021makes,agrawal2022large,huang2022language}. To retrieve the most relevant examples, we employ a retrieval model $\theta \in \Theta$, which selects top-$p$ similar sentences as evidences based on lexical or semantic similarity. Formally,
% \begin{equation}
% \phi_{\mathrm{LLM}}(\vec{x}, \Tilde{\vec{x}}_j,\mathcal{E}_{p;\theta}(\vec{x})) \mapsto \vec{x}'_j,
% \label{eq:rag}
% \end{equation} 
% where, $\mathcal{E}_{p;\theta}(\vec{x}) \subset \mathcal{T}$ and $\vec{x}'_j$ is the predicted corrected sentence of the masked sentence $\Tilde{\vec{x}}_j$ generated by an LLM. For details regarding our prompt template, please refer to Figure \ref{fig:m2c} of Appendix \ref{tempdix}.

Given the set of masked claims $\{\tilde{\vec{x}}_j\}_{j=1}^k$ derived from an input claim $\vec{x}$, this module performs evidence guided correction through a RAG based approach. Intuitively, the masking step isolates uncertain spans, but without external grounding, directly reconstructing them risks hallucination or semantic drift. To mitigate this, we shift from purely generative reconstruction to an evidence based generation process, where corrections are explicitly anchored in retrieved knowledge.  Here, the model directly leverages a small set of examples which are similar to the current claim from an external corpus $\mathcal{T}$, eventually including these with the masked claim as a part of an input prompt to an LLM~\cite{liu2021makes,agrawal2022large,huang2022language}. To retrieve the most relevant examples, we employ a retrieval model $\theta \in \Theta$, which selects top-$p$ similar sentences as evidences based on lexical or semantic similarity. Formally,
\begin{equation} \phi_{\mathrm{LLM}}(\vec{x}, \Tilde{\vec{x}}_j,\mathcal{E}_{p;\theta}(\vec{x})) \mapsto \vec{x}'_j, \label{eq:rag} \end{equation}
where, $\mathcal{E}_{p;\theta}(\vec{x}) \subset \mathcal{T}$ and $\vec{x}'_j$ is the predicted corrected sentence of the masked sentence $\Tilde{\vec{x}}_j$ generated by an LLM. For details regarding our prompt template, please refer to Figure \ref{fig:m2c} of Appendix \ref{appx:prompt_temp}.

\subsection{Correction Scoring} \label{sec:corr_score}
In this module, given a set of generated candidate corrected sentences $\{\vec{x}'_j\}_{j=1}^k$, appended with input claim $\vec{x}$, our objective is to identify the most factually consistent and semantically faithful candidate correction $\hat{\vec{x}}$ for a given input claim $\vec{x}$. We utilize a scoring function $\mathcal{F}(\vec{x}'_j)$ by using two metrics as used in \cite{huang2023zero}: 1) DocNLI~\cite{yin-etal-2021-docnli}, which measures factual entailment with the provided evidence $\mathcal{E}_{p;\theta}(\vec{x})$, and 2) ROUGE-L~\citep{lin2004rouge}, which captures the longest common subsequences between the input claim $\vec{x}$ and the candidate $\vec{x}'_j$, thereby ensuring minimal deviation from the input claim. The candidate having the maximum score is picked as the final corrected claim, $\hat{\vec{x}}$. Formally speaking, $\hat{\vec{x}}= \arg\max_{\vec{x}'_j} \mathcal{F}(\vec{x}'_j)$,  where,
% \begin{align*}
% \small
% \hat{\vec{x}} &= \arg\max_{\vec{x}'_m} \mathcal{F}(\vec{x}'_m)\\
% \text{where,}\\
% \mathcal{F}(\vec{x}'_m)
% &=\lambda \cdot \text{DocNLI}(\vec{x}'_m, \mathcal{E}) 
% + (1 - \lambda) \cdot \text{ROUGE-L}(\vec{x}, \vec{x}'_m)
% \end{align*}

% \begin{align*}
% \hat{\vec{x}} &= \arg\max_{\vec{x}'_m} \mathcal{F}(\vec{x}'_m)\\
% \text{where,}\quad
% \mathcal{F}(\vec{x}'_m)
% &= \lambda \cdot \text{DocNLI}(\vec{x}'_m, \mathcal{E}) \\
% &\quad + (1 - \lambda) \cdot \text{ROUGE-L}(\vec{x}, \vec{x}'_m)
% \end{align*}

% {\small
% \begin{align*}
% \hat{\vec{x}} &= \arg\max_{\vec{x}'_m} \mathcal{F}(\vec{x}'_m)\\
% \text{where,}\quad
% \mathcal{F}(\vec{x}'_m)
% &= \lambda \cdot \text{DocNLI}(\vec{x}'_m, \mathcal{E}) \\
% &\quad + (1 - \lambda) \cdot \text{ROUGE-L}(\vec{x}, \vec{x}'_m)
% \end{align*}
% }
\vspace{-5mm}
\begin{equation*}
\small
\mathcal{F}(\vec{x}'_j)
= \lambda \cdot \text{DocNLI}(\vec{x}'_j, \mathcal{E})+ (1 - \lambda) \cdot \text{ROUGE-L}(\vec{x}, \vec{x}'_j).
\end{equation*}
%
% \vspace{-6mm}
% \begin{equation*}
% \scriptsize
% \mathcal{F}(\vec{x}'_j)
% = \lambda \cdot \text{DocNLI}(\vec{x}'_j, \mathcal{E})+ (1 - \lambda) \cdot \text{ROUGE-L}(\vec{x}, \vec{x}'_j)
% \end{equation*}
% \vspace{-6mm}
% \begin{align}
% \hat{\vec{x}} &= \arg\max_{\vec{x}'_j} \mathcal{F}(\vec{x}'_j)
% \end{align}
% \vspace{-1mm}
The parameter $\lambda \in [0, 1]$ balances factuality and faithfulness. We selected the optimal $\lambda$ using FEVER validation data (see Appendix \ref{appx:lambda}). The schematic diagram of M$_2$C is depicted in Figure \ref{fig:archit}.
%////////
% As illustrated in Figure \ref{fig:archit}, in this framework, for each input claim, we first identify and mask tokens that are likely to be incorrect. To bypass the use of gold evidence, relevant sentences are retrieved using IR models from an external corpus. Given the masked sentences, original input, and retrieved evidence, an LLM generates candidate corrections. Finally, a scoring function is used to select the final output by minimizing unnecessary changes to the input claim.
%//////
% {\small
% \begin{align}
% \mathcal{F}(\vec{x}'_m)
% &= \lambda \cdot \text{DocNLI}(\vec{x}'_m, \mathcal{E})+ (1 - \lambda) \cdot \text{ROUGE-L}(\vec{x}, \vec{x}'_m) \tag{1} \label{eq:scoring_function}
% \end{align}
% \begin{align*}
% \hat{\vec{x}} &= \arg\max_{\vec{x}'_m} \mathcal{F}(\vec{x}'_m)
% \end{align*}
% }

\subsection{Ensemble Based Model}
% We propose a novel ensemble model that aggregates the final corrected predictions generated using top-retrieved documents obtained from multiple ranking models to enhance the reliability and robustness of our system. For a given masked input, with each retriever $\theta_i \in \Theta$ (where $i=1, \ldots, m$) M$_2$C  generates a candidate correction. Here we employ the majority voting principle based on the hard voting strategy to select the final corrected claim, i.e., the candidate that receives maximum votes across all retrievers is chosen as the final output for the downstream task (see Figure \ref{fig:archit}). \PS{To the best of our knowledge, this is the first work to apply majority voting at the correction level in fact correction, where each retriever independently generates a complete corrected claim. This design mitigates retriever-specific biases while avoiding the need for additional annotations or model training.}
% We name this method \underline{M}ask-\underline{to}-\underline{C}orrect$^+$ (M$_2$C$^+$)\footnote{The code is publicly available at: \url{https://github.com/payelsantra/MaskToCorrect.git}}
% Code available at: \url{https://anonymous.4open.science/r/MaskToCorrect-5B5F/}}.

We further extend our framework with a novel ensemble mechanism that aggregates corrected outputs derived from multiple retrieval perspectives to improve robustness and reliability. Intuitively, while a single retriever may capture only a partial or biased view of the evidence space, different retrievers $\theta_i \in \Theta$ (for $i=1,\ldots,m$) often surface complementary contexts, leading to diverse yet plausible corrections for a given masked input. Instead of committing to any single retrieval signal, we shift to a consensus driven paradigm, where M$_2$C generates a set of candidate corrections, one per retriever and consolidates them via a hard majority voting strategy. Concretely, the final corrected claim is selected as the candidate receiving the highest number of votes across all retrievers, effectively approximating a form of collective agreement over independently grounded generations (see Figure~\ref{fig:archit}). This design mitigates retriever-specific biases, enhances stability under retrieval noise, and eliminates the need for additional supervision or training. We named this  \underline{M}ask-\underline{to}-\underline{C}orrect$^+$ or, M$_2$C$^+$~\footnote{The code is publicly available at: \url{https://github.com/payelsantra/MaskToCorrect.git}}.

\section{Experiment Setup}
% \subsection{Evaluation}

\subsection{Research Questions}
The effectiveness of RAG depends on both the quality of the retrieval and the masking strategy used. To understand their impact on downstream performance, we investigate the following research questions:
% We investigate the following research questions: 
a)~ \textbf{RQ-1}: Does M$_2$C$^+$ perform better than the individual retriever-based approach, M$_2$C? b) \textbf{RQ-2}: Can diversity-aware masking strategies lead to improved downstream performance? c) \textbf{RQ-3}: How does performance with retrieved evidence differ compared to the gold-standard evidence? d) \textbf{RQ-4}: How strongly does retrieval effectiveness correlate with downstream gains?
% \begin{itemize}
    % \para{RQ-1} Does M$_2$C$^+$ perform better than the individual retriever-based approach, M$_2$C? 
    % \para{RQ-2} Can diversity-aware masking strategies lead to improved downstream performance?
    %  \para{RQ-3} How does performance with retrieved evidence differ from that with gold-standard evidence?
    %  \para{RQ-4} How strongly does retrieval effectiveness correlate with downstream gains?
% \end{itemize}

% \begin{itemize}
%     \item[] \textbf{RQ-1}: Does ensembling multiple retriever-based corrections improve performance over individual retrievers?
%     \item[] \textbf{RQ-2}: Can diversity-aware masking strategies lead to improved downstream performance?
%     \item[] \textbf{RQ-3}: How does performance with retrieved evidence differ from that with gold-standard evidence?
% \item[] \textbf{RQ-4}: How strongly does retrieval effectiveness correlate with downstream gains?
% \end{itemize}

\subsection{Dataset Description}
We conduct our experiments on two datasets repurposed for our task: 1) FEVER~\cite{thorne-vlachos-2021-evidence}, comprising of general-domain claims 2) SciFact~\cite{wadden-etal-2020-fact}, consisting of scientific claims. For FEVER we retrieve relevant evidence from the Wikipedia'18 dump, and for SciFact we use the S2ORC corpus \cite{wadden-etal-2020-fact}. Table~\ref{tbl:datasets_statistics} summarizes these dataset statistics. Detailed description can be found in Appendix~\ref{appx:data_desc}.

\begin{table}[h]
\centering
\small
\begin{adjustbox}{width=.9\columnwidth}
\begin{tabular}{@{}llccr@{}}
\toprule
Dataset &
Usage &
\#Sup & \#Ref & \#Claims \\ 
\toprule
 % FEVER-sent
  \multirow{2}{*}{FEVER} &
 Validation &414 &601 & 1555 \\ 
 \cmidrule{2-5}
  & \multirow{2}{*}{Test} & 1,593 & 2,289 & 3,882\\ 
  % Sci-Fact-sent
 SciFact &  & 43 & 57 & 100 \\ 
\bottomrule
\end{tabular}%
\end{adjustbox}
\caption{Statistics of FEVER and SciFact dataset for our downstream fact-correction task.}  
% The top part corresponds to large datasets, and the bottom to small datasets.
\label{tbl:datasets_statistics}
\end{table}

\subsection{Rankers Investigated}
In this work, we employ RAG in the evidence-guided corrector module and analyze the impact of context selection. We experiment with four single-stage retrievers and one multi-stage ranker. We consider a diverse set of widely used, training-free retrievers and rerankers to mitigate retriever-specific biases and ensure a fair comparison. For consistency, we retrieve the top 50 candidates for each claim across all models.

\textbf{Single-stage Ranker.}
We explore two \textit{sparse retrievers}: 1) \textbf{BM25}~\cite{10.1561/1500000019}, a traditional lexical retriever that relies on exact term matching, utilizing a TF-IDF variant. In our experiments, we set the BM25 parameters $k_1$ and $b$ to $0.9$ and $0.4$, respectively. 2) \textbf{RM3}~\cite{10.1145/3471158.3472261}, a pseudo-relevance feedback method (built on top of lexical retrievers like BM25) to expand the queries and re-rank the retrieved documents.

We also experimented with two \textit{dense end-to-end rankers}: 1) \textbf{Contriever}~\cite{izacard2021unsupervised}, a bi-encoder framework fine-tuned on the MS MARCO passage dataset to generate dense embeddings and capture semantic similarity. 2) \textbf{ColBERT}~\cite{10.1145/3397271.3401075}, adopts TCT-ColBERT, a late interaction model, to efficiently capture fine-grained token-level interactions.

% \uls
% \li \textbf{Sparse Ranker} 
% \para{BM25}~\cite{10.1561/1500000019}: BM25 is a traditional \textit{lexical} retriever that generally relies on exact term matching, utilizing a TF-IDF variant. In our experiments, we set the BM25 parameters $k_1$ and $b$ to $0.9$ and $0.4$, respectively for candidate retrieval.

% \para{RM3}~\cite{10.1145/3471158.3472261}: An extension of BM25 that uses pseudo-relevance feedback to expand queries with additional terms from the top-ranked documents, improving recall.

% \li \textbf{Dense Ranker}
%  \para{Contriever}~\cite{izacard2021unsupervised}: In our experiment, we utilize a \textit{dense end-to-end} model, Contriever\footnote{\url{https://huggingface.co/facebook/contriever-msmarco}}, a bi-encoder framework fine-tuned on the MS MARCO passage dataset to generate dense embeddings and capture semantic similarity. 

% \para{ColBERT}~\cite{10.1145/3397271.3401075}: 
% For this experiment adopt the TCT-ColBERT model\footnote{\url{https://huggingface.co/castorini/tct_colbert-v2-hnp-msmarco}}, which is a \textit{dense end-to-end} model distilled from ColBERT, designed to efficiently capture fine-grained token-level interactions. 

\textbf{Two-stage Ranker.} In our \textit{retrieve-and-rerank} approach, we utilize \textbf{MonoT5}~\cite{pradeep2021expando}, which uses BM25 to retrieve a small candidate set, which is then refined by a T5-based MonoT5 (a cross-encoder model) re-ranker. 

In this paper, our retriever selection covers lexical, pseudo-relevance feedback, dense (bi-encoder, late-interaction), and cross-encoder reranker paradigms. By relying mostly on CPU-compatible methods, our framework remains resource-efficient while enabling robust analysis of retriever variability.

% Recent studies show, when a rank list is re-ranked and increase semantic diversity often helps in downstream tasks~\cite{carraro2024enhancing}. 

% we first retrieve the top $100$ candidate documents using the same configurations as our single-stage retriever, and then re-ranks (also configured identically as explained in single-stage retriever) to narrow these down to the top $50$ documents. 
 
% \textbf{MonoT5}~\cite{pradeep2021expando}: This belong to retrieve-and-rerank pipeline, which uses BM25 to retrieve a small candidate set, which is then refined by a T5-based MonoT5\footnote{\url{https://huggingface.co/castorini/monot5-base-msmarco}}(a cross-encoder model) re-ranker. 
% \ule

\subsection{Maskers Investigated}
We compare our proposed masking strategy with the following baseline maskers: \\
(a) \textbf{Random Masking (RM)}~\cite{thorneEvi}: Randomly masks tokens without considering their contextual relevance. (b) \textbf{Heuristic Masking (HM)}~\cite{thorneEvi}: Masks only the tokens that are present in the claim but absent from the evidence (based on heuristics, these are more likely to be factually incorrect). (c) \textbf{Entity Masking (EM)}~\citep{huang2023zero}: It utilizes the entity set $E_s(\vec{x})$ (introduced in Section~\ref{sec:masker}) to mask entities/phrases extracted from the input claim. For the claim mentioned in Figure \ref{fig:archit}, EM provides entities like `Pulmonary embolism', `high blood oxygen levels', `Pulmonary' etc to mask. But this strategy does not consider redundancy or semantic importance of the phrases. Moreover, sometimes may mistakenly mask irrelevant tokens like ``oxygen”, ``indicated”, etc. Thus, relying only on entities may affect the model's performance.

Our proposed \textbf{Diversity Masking (DM)}: It employs an MMR–based selection, which is a greedy algorithm and re-ranks entities in $E_s(\vec{x})$ to maximize their similarity to the claim while minimizing similarity between the selected tokens, thereby enhancing EM by ensuring both semantic relevance and diversity, as detailed in Section~\ref{sec:masker}.

\subsection{Methods Investigated} We compare our proposed methodology with the following baselines, considering two settings for each: 1) using evidence retrieved from an external corpus, and 2) using gold-standard evidence.

\textbf{Non-parametric Baselines.} These methods do not involve any parametric training for our downstream fact correction task. We employ the following: a) \textbf{\iclnz}~\cite{labruna2024retrieve,kojima2022large,li2023classification}: To assess the importance of masking, in this approach, given a claim, the model directly prompts to generate the corrected claim without utilizing any example. 
% Here the model relies on its internal capabilities.
%
% To assess the role of masking, this approach prompts the model to generate a corrected version of the claim directly, without using any examples. The model relies solely on its internal knowledge to perform the correction.
%
b) \textbf{RAG}~\cite{labruna2024retrieve}: Similar to the previous approach, here we directly prompt the LLM to correct a given claim, but in the presence of some context. 
% \li \textbf{M$_2$C$_\text{W/oE}$}: This baseline investigates the need to utilize evidences for our proposed M$_2$C pipeline. Specifically, we evaluate the performance of M$_2$C when the evidence-guided corrector module is in zero-shot setting (i.e., \underline{W}ith\underline{o}ut using any \underline{E}vidence). This allows us to explore the effectiveness of evidence on the overall correction quality.
%
c) \textbf{M$_2$C$_\text{w/Ver}$}: This variant incorporates an initial verification step in the M$_2$C framework by using an LLM, with corrections applied only to claims identified as incorrect.
d) \textbf{Z{\tiny ERO}F{\tiny EC}-DA}~\cite{huang2023zero}, a training-free QA-based framework that divides the task into five sub-tasks. Here, RoBERTa~\cite{liu2019roberta} was fine-tuned on two biomedical datasets to improve domain-specific performance. 
\begin{table*}[t]
\centering
\small
\begin{adjustbox}{width=0.8\textwidth}
\begin{tabular}{@{}lll rrrrrrrr@{}}
\toprule
 & & &\multicolumn{4}{c}{FEVER} & \multicolumn{4}{c}{SciFact}\\
\cmidrule(l){4-7}
\cmidrule(l){8-11}
 \multicolumn{2}{c}{}& & \multicolumn{2}{c}{SARI (\%)}& \multicolumn{2}{c}{BART} & \multicolumn{2}{c}{SARI (\%)}& \multicolumn{2}{c}{BART}  \\
\cmidrule(r){4-5}
\cmidrule(r){6-7}
\cmidrule(r){8-9}
\cmidrule(r){10-11}
  LLM &Predictor & Ranker  & Retrieved & Gold  & Retrieved & Gold & Retrieved & Gold & Retrieved & Gold \\
\midrule
\multirow{3}{*}{n/a}   &Z{\tiny ERO}F{\tiny EC}-DA  & MonoT5 & 39.3734 & 40.7713 & -2.8629& -2.7623 & 32.1194 & 32.1988  & -3.1905 & -3.1756\\
 & T5-Distant & GENRE &  34.9734&  43.7252& -5.4120&-2.4015 & 23.3640&22.9304  &-3.0965&  -2.9803\\
  & C{\tiny OMP}E{\tiny DIT} & MonoT5 &  26.0694 &30.8499
  & -3.1801 & -2.9100 & 30.4219& 30.6196 & -3.1699& -3.0591\\
% && Vence &     &  & & & &  &&  \\
%  & &LIFE &     &  & & & &  && \\
\midrule
%  &  &  &     &  &  &   & &  &  &  &  &\\
%  \midrule
 \multirow{10}{*}{\rotatebox{90}{Llama}} & 0-shot & n/a &  \multicolumn{2}{c}{44.9646}  & \multicolumn{2}{c}{-3.7971} & \multicolumn{2}{c}{35.0145} & \multicolumn{2}{c}{-2.9990}\\
   % &  & M$_2$C-DM$_{\text{W/oE}}$     &  n/a  & \multicolumn{2}{c}{} &  \multicolumn{2}{c}{} & \multicolumn{2}{c}{}  &  \multicolumn{2}{c}{}  \\ 
  & RAG    &   MonoT5  & 41.7222 &   42.7340 &  -2.3939& -2.0769&  34.3641&36.5339  & -2.6404  & -2.3658 \\
  % \rowcolor{yellow}
   & M$_2$C$_\text{w/Ver}$    &   MonoT5  & 48.2327 & 49.5799  & -2.6143 & -2.4591 &34.2985   & 30.6697 &-2.9546  & -3.1277 \\
    \cmidrule(r){2-11}
    & \multirow{5}{*}{M$_2$C$_\text{DM}$}  &   BM25    & 45.0705 & \multirow{6}{*}{\textbf{51.2129}} & -2.5935   & \multirow{6}{*}{\underline{-2.3720}} & 34.8178 & \multirow{6}{*}{\textbf{40.0456}} &  -2.9862 &   \multirow{6}{*}{\textbf{-2.6892}} \\ 
    &  & RM3 & 45.5321 &  & -2.6088  & & 36.8477 & & -2.9852 &  \\ 
   &  &  Contriever & 48.6495 &  &  -2.6125 & & 37.1026 &  & -2.9409 &  \\ 
      & & ColBERT & 47.4645 &  & -2.6491  & & 36.4724 & & -2.9212 &   \\ 
  &  & MonoT5 & 49.1997 & &  -2.6702 & & \underline{39.6940} & & -2.9304&  \\ 
  % \rowcolor{yellow}
  & \gc{M$_2$C$^+_\text{DM}$}     & \gc{n/a}  & \gc{\underline{49.3749}}	 &  \gc{} & \gc{\textbf{-2.5939}}   &\gc{} &  \gc{37.5820} & \gc{} & \gc{\underline{-2.9302}} &  \gc{} \\
\cmidrule(r){1-11}
 \multirow{10}{*}{\rotatebox{90}{Qwen}} & 0-shot   & n/a &  \multicolumn{2}{c}{43.2337} & \multicolumn{2}{c}{-2.7315} & \multicolumn{2}{c}{35.7045} & \multicolumn{2}{c}{-3.0843}\\
 % &  & M$_2$C-DM$_{\text{W/oE}}$         & n/a &  &   & &  &  &  &  \\
 & RAG  &  MonoT5   & 46.9218  &   48.5516&-2.7590&-2.5020  & 36.3252  &34.7780& -2.4655&  -2.8676 \\
 % \rowcolor{yellow}
 & M$_2$C$_\text{w/Ver}$    &   MonoT5  &   49.8118	& 51.0140  & -3.0608& -2.4595& 40.9380	 & 42.6284 & -2.7091 & -2.6048 \\
  \cmidrule(r){2-11}
  & \multirow{5}{*}{M$_2$C$_\text{DM}$}  &   BM25  &45.9969 &  \multirow{6}{*}{\textbf{51.2783}} & -2.7278  & \multirow{6}{*}{\textbf{-2.3908}} & 35.6633 &   \multirow{6}{*}{\textbf{42.6331}}& -2.8656 &   \multirow{6}{*}{\textbf{-2.1453}} \\ 
  &  & RM3 & 46.5141&  &  -2.6143 & & 39.1719 &   &-2.9198 &  \\ 
  &  & Contriever & 49.2850&  &  -2.6159 & & 37.1282 &  & -2.9282 &  \\ 
  & & ColBERT &48.3629 &  &   -2.6326& &37.7490  &  & -2.9198  & \\ 
  &  & MonoT5 &50.3638 &  & -2.6848  & & \underline{41.7093} & & \underline{-2.8450} &  \\ 
  % \rowcolor{yellow}
   &  \gc{M$_2$C$^+_\text{DM}$}   & \gc{n/a} &  \gc{\underline{50.7141}} &  \gc{} &  \gc{\underline{-2.5724}} & \gc{} & \gc{38.4647} & \gc{} & \gc{-2.8636} & \gc{}  \\
\bottomrule
\end{tabular}
\end{adjustbox}
\caption{\small Performance of M$_2$C$^+$ and M$_2$C in diversity-aware entity masking setting (i.e., `DM' subscript) relative to the baselines. For each LLM, bold letters indicate the best performance among all models with retrieved/gold evidence, and the second-best results are underlined. In the table, all the RAG-based approaches (i.e., RAG,  M$_2$C and M$_2$C$^+$) were obtained with a context size of 3, i.e., $p=3$ in Equation~\ref{eq:rag}. The results of M$_2$C$^+$ are in gray. Since MonoT5 performed best for M$_2$C setting, we chose MonoT5 for all the retrieval-based baseline experiments.}
\vspace{-1em}
\label{tab:combined_results}
\end{table*}

% \input{table/table_5ret}
% \ule

\textbf{Distantly Supervised Baselines.} We employ the following parametric approaches as baselines.
a) \textbf{T5-Distant}~\cite{thorne2021evidence}: Similar to our methodology, it also follows a mask-then-correct approach using DPR \cite{karpukhin2020dense} and GENRE \cite{de2020autoregressive} for evidence retrieval. It utilizes a heuristic masker and a finetuned T5-base corrector. 
% this paper also utilizes mask-and-correct strategy. 
% , achieving notable improvements in human evaluation and SARI score without requiring gold correction labels.
%
b) \textbf{C{\tiny OMP}E{\tiny DIT}:}~\cite{fabbri2022improving}: This method enhances summary factuality through post-editing. It uses BART model trained on sentence compression data to revise summaries by removing entities absent from the source document. These entities are detected using NER and highlighted with special tokens.

% We include this method to enhance summary factuality through post-editing. It uses a BART model trained on sentence compression data to revise summaries by removing entities absent from the source document. These entities are detected using named entity recognition and highlighted with special tokens.
%We include this method to evaluate hallucination mitigation through post-editing. It employs a BART model trained on sentence compression data to remove extrinsic entity errors by editing summaries marked with special tokens, based on the source document.
% It employs a BART-large model trained on sentence compression data to remove extrinsic entity errors. The perturber marks hallucinated entities, and the editor generates factually consistent summaries by conditioning on the source and perturbed input, ensuring improved entity precision without sacrificing informativeness or fluency.
% \ule

\para{Our Proposed Approaches} We experiment with two variants of our framework: a) \textbf{M$_2$C}, which employs individual rankers within the three-stage pipeline for selecting examples, and b) \textbf{M$_2$C$^+$}, which ensembles corrected outputs from multiple rankers to mitigate noise and increase reliability. 

% We explored two different variants of our proposed M$_2$C-based methodology. 1) M$_2$C, where we apply our three-stage pipeline using individual rankers for example selection and 2) M$_2$C$^+$, where we used ensemble technique to combine the corrected claims from each rankers and mitigate noise from individual ranker decisions.

% \subsection{/LLM Settings and Evaluation}
\subsection{Evaluation Metrics}
We use BART score~\cite{yuan2021bartscore} and SARI~\cite{lewis2019bart} as our evaluation metrics. BART score measures semantic similarity between the generated prediction and the ground-truth evidence. SARI score helps to evaluate the quality of edits made by the corrector. In both cases, higher scores indicate more accurate and faithful corrections. A detailed description is provided in Appendix~\ref{appx:metrics}. Additionally, to quantify the retrieval effectiveness of each IR model, we report nDCG@10 metric.
%
% We evaluate correction quality using BARTScore and SARI. BARTScore~\cite{yuan2021bartscore} assesses the semantic alignment between the generated claim and the evidence, while SARI~\cite{xuoptimizing} measures the quality of edits by comparing the generated output against the original and reference claims. Higher values in both indicate better factual accuracy and faithfulness. To assess retrieval performance, we also report nDCG@10, which captures the ranking quality of retrieved evidence.

\subsection{Implementation Details} 
We conduct all our experiments using the LLM Llama-2.0 (70B)~\citep{touvron2023llama} and Qwen-2.5 (32B)~\citep{qwen2}, chosen
after preliminary experiments with various open-source LLMs like LLaMA-3, Qwen-2, and Mistral etc. Notably, these models belong to two distinct decoder-only families, enabling evaluation across different architectures. For diversity-aware masking, we applied $\alpha$ as $0.3$ and the number of masked sentences $m$ as 10 (see Section \ref{sec:sensitivity_anal}) and used Pyserini~\cite{lin2021pyserini} for retrieval. Further details have been provided in Appendix \ref{appx:exp_env}. Our ensemble framework uses the four best-performing retrievers (excluding RM3) identified through preliminary analysis (discussed in Section \ref{sec:sensitivity_ret}).

\begin{figure*}[t]%
    %\vspace{-.5cm}
    \centering
    \subfloat[
    %\centering
    %PwC KB Method Components Page
    \small Sensitivity on FEVER
    ]{{\includegraphics[width=.65\columnwidth]{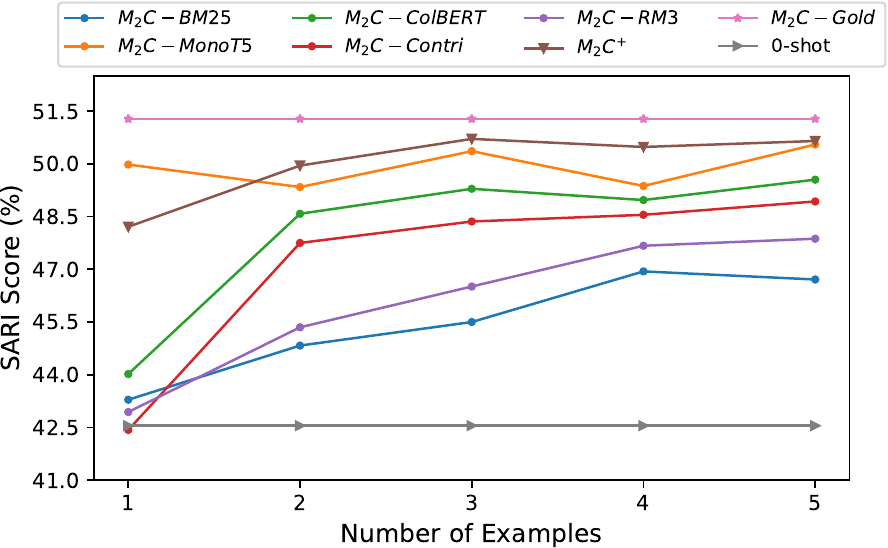}}
    %sensi_fever_0shot_crop.pdf
    \label{fig:fever_sensitivity}
    }%
    % \qquad
    \subfloat[
    \centering
    %PwC KB Method Components Paper Page
    \small Sensitivity on SciFact
    ]
    {{\includegraphics[width=.65\columnwidth]{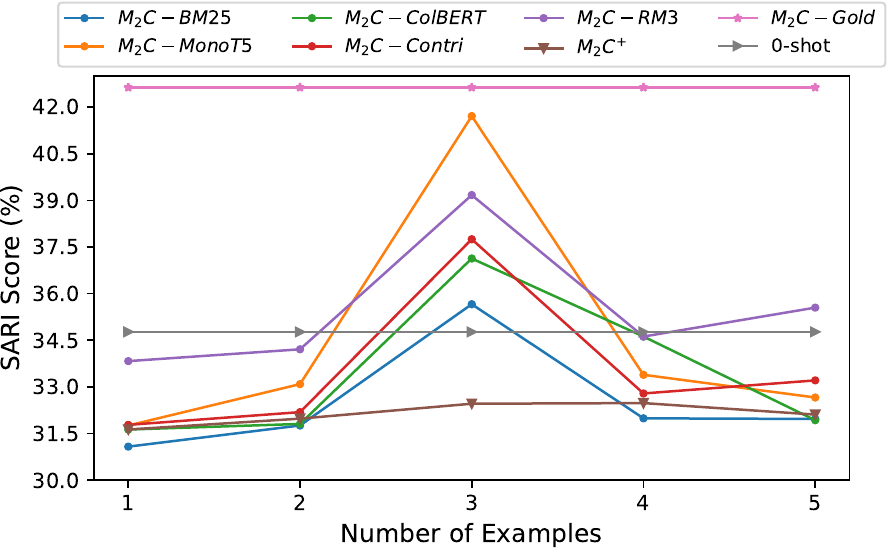}} %figure/sensi_scifact_output1.pdf
    \label{fig:sci_sensitivity}
    }%\
    % \hfill
    \subfloat[
    \centering
    %PwC KB Method Components Paper Page
    \small Comparing Correction Scoring
    ]
    {{\includegraphics[width=.35\columnwidth]{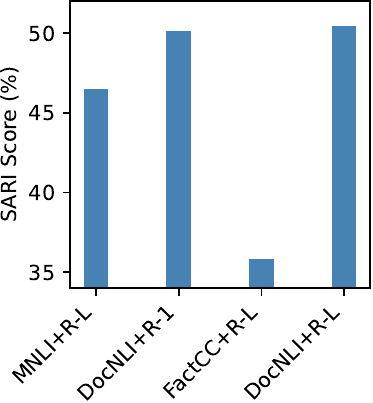}}
    \label{fig:score_func}
    }%
% \hfill
    \subfloat[
    \centering
    %PwC KB Method Components Paper Page
    \small Sensitivity of M$_2$C$_{\text{RM}}$ on FEVER
    ]
    {{\includegraphics[width=.35\columnwidth]{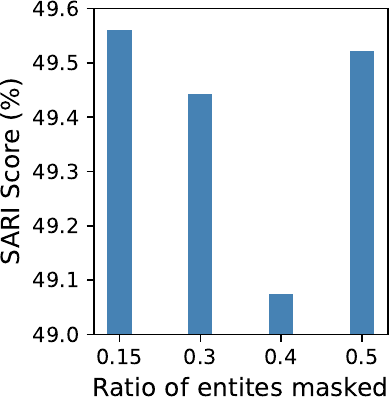}}
    \label{fig:rm_sensitivity}
    }%
    \\
    \subfloat[
    \centering
    %PwC KB Method Components Paper Page
    \small Correlation on FEVER
    ]
    {{\includegraphics[width=.39\columnwidth]{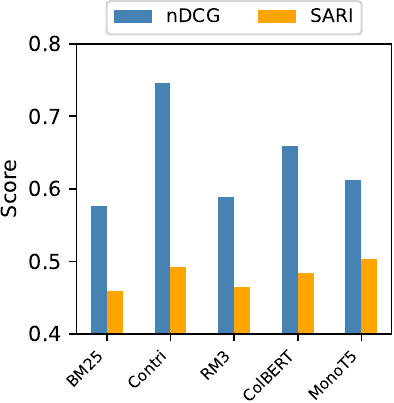} }
    \label{fig:correlation}
    }%
    \hfill
    \subfloat[
    %\centering
    %PwC KB Method Components Page
    \small Comparison of four maskers on FEVER
    ]{{\includegraphics[width=.36\columnwidth]{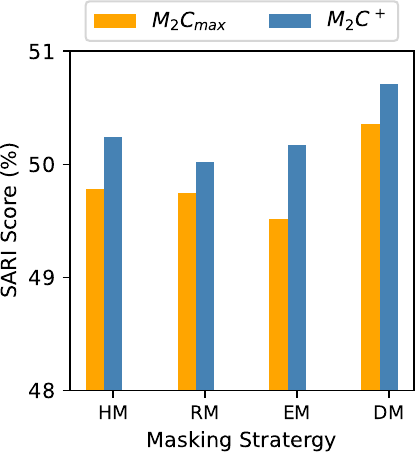}} 
    \label{fig:fever_mask}
    }%
    \hfill
    \subfloat[
    %\centering
    %PwC KB Method Components Page
    \small Comparison of four maskers on SciFact
    ]{{\includegraphics[width=.36\columnwidth]{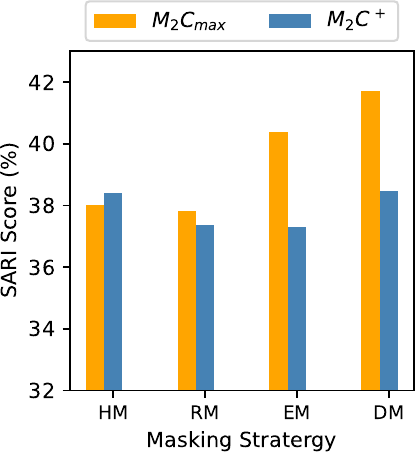}} 
    \label{fig:sci_mask}
    }%
    \hfill
    \subfloat[
    %\centering
    %PwC KB Method Components Page
    \small Sensitivity of $\alpha$ of M$_2$C$_\text{DM}$ on FEVER
    ]{{\includegraphics[width=.36\columnwidth]{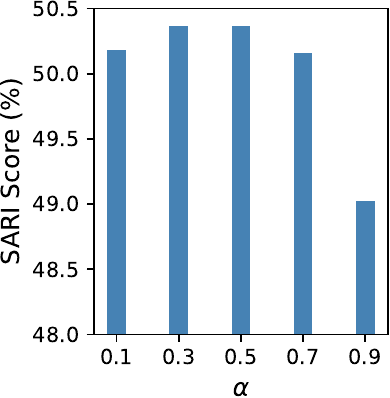}} 
    % [50.1850, 50.3638, 50.3629, 50.1553, 49.02]
    \label{fig:DM_alpha}
    }%
    \hfill
    \subfloat[
    %\centering
    %PwC KB Method Components Page
    \small Sensitivity of m of M$_2$C$_\text{DM}$ on FEVER
    ]{{\includegraphics[width=.34\columnwidth]{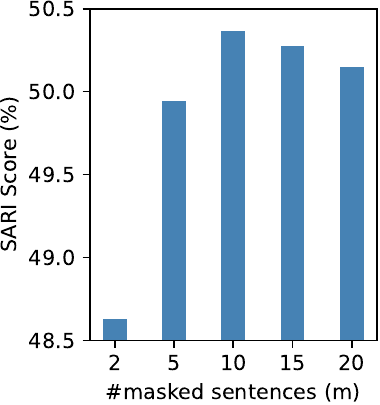}} 
    \label{fig:DM_sent}
    }%
    %\vspace{-0.3cm}
%     \caption{\small Sensitivity of the proposed M$_2$C model variants with number of in-context examples on (a) FEVER, (b) SciFact Dataset using Qwen model. M$_2$C-Gold denotes M$_2$C with gold annotated evidence. (c) Correlation between retrieval quality (nDCG@10) and downstream task performance (SARI) for M$_2$C$_\text{DM}$ model using Qwen. For fair comparison, SARI scores (\%) have been normalized by dividing them by $100$. (d) \textcolor{red}{RM sensitivity, ratio chnge-FEVER- QWEN- MONOT5 -- 0.3,0.5,0.05,0.15,0.37}.. (f,g) Comparison of various masking strategies, e.g., Random (RM), Heuristic (HM),
% Entity (EM), and the proposed Diversity-aware Masking (DM) for both M$_2$C variants using Qwen as the base model. M$_2$C$_\text{max}$ signifies the best performance of M$_2$C across all retrievers. \textcolor{red}{(h) Sensitivity of $\alpha$ on DM (h) Sensitivity of number of seleted sentences in DM.,,, (e)Comparison of different metric combinations for the correction scoring module on the M$_2$C$^+_\text{DM}$ (MonoT5) model for the FEVER test set using Qwen.}}

\caption{\small Sensitivity of M$_2$C variants using the Qwen model. (a–b) Sensitivity to the number of in-context examples on FEVER and SciFact; M$_2$C-Gold denotes use of gold-standard evidence. (c) Effect of different correction scoring combinations on M$_2$C$_\text{DM}$–MonoT5 for FEVER. (d) Sensitivity of token masking ratio of M$_2$C$_\text{RM}$–MonoT5 on FEVER. (e) Correlation between retrieval quality (nDCG@10) and downstream performance (SARI(\%),  normalized by dividing them by 100) for M$_2$C$_\text{DM}$. (f–g) Comparison of masking strategies for M$_2$C variants on FEVER and SciFact; M$_2$C$_\text{max}$ indicates the best performance of M$_2$C across all retrievers. (h–i) Effect of diversity parameter $\alpha$ and number of masked sentences ($m$) in M$_2$C$_\text{DM}$–MonoT5 on FEVER.}

    \vspace{-1em}
    \label{fig:3_figs}
\end{figure*}
\section{Result and Analysis}

% Table \ref{tab:combined_results} shows a comparison between our proposed approaches and several baselines for two benchmark datasets using two different LLMs (Llama and Qwen). 
\subsection{Main Observations}
To address \textbf{RQ-1} \textbf{(Performance of proposed methodology)}, Table \ref{tab:combined_results} shows that both our proposed variants outperform all parametric and non-parametric baselines for each dataset, across both LLMs. Notably, our method outperforms the verification-based variant M$_2$C$_\text{w/Ver}$, indicating that the multi-stage correction pipeline already ensures strong factual alignment.
%
% the likely reason is that introducing a separate verification stage does not add complementary information—our multi-stage correction pipeline already captures the necessary factual alignment.
On FEVER, M$_2$C$^+_\text{DM}$ performs better than M$_2$C$_\text{DM}$ (shown in Figure \ref{fig:fever_mask}). In contrast, on SciFact, a dataset of scientific domain, M$_2$C$_\text{DM}$ with MonoT5 performs better than M$_2$C$^+_\text{DM}$ (shown in Figure \ref{fig:sci_mask}). The likely reason is that in specialized domains, a single high-performing retriever (like MonoT5, in Table \ref{tab:combined_results}) may already retrieve sufficiently relevant evidence, whereas ensembling may lead to the selection of non-relevant evidence retrieved by multiple underperforming retrievers, thereby degrading the performance. Moreover, the effectiveness of our model in real-world scenarios is discussed in Appendix~\ref{appx:real_WORLD}.

In relation to \textbf{RQ-2} \textbf{(Choice of masker)}, in this work, we experiment with four masking strategies: Random (RM), Heuristic (HM), Entity (EM), and our proposed Diversity-aware Masking (DM). Figures~\ref{fig:fever_mask} and \ref{fig:sci_mask} compare the performance (SARI score (\%)) of M$_2$C$_\text{max}$ (reports the best performance of M$_2$C across five retrievers) and M$_2$C$^+$ models across these masking strategies. We observe that, for both datasets, DM and HM generally outperform RM and EM in both M$_2$C variants, although DM, EM outperform HM and RM on the SciFact dataset under the M$_2$C$_\text{max}$ setting. In FEVER, which is a general domain dataset, DM significantly outperforms HM for both  M$_2$C variants. The likely reason is that MMR effectively balances semantic relevance and diversity, allowing it to handle noisy or paraphrased evidence. In contrast, HM strictly relies on token-level overlaps, which leads to false negatives. On the other hand, in the SciFact dataset, collected from the scientific domain, HM’s direct token matching is more effective in scientific texts. Moreover, the limited linguistic variability in SciFact reduces the benefit of selecting diversity-based entities. Thus, DM achieves comparable result to HM for both models.

% \input{fig_def/error}
% In relation to \textbf{RQ-3} \textbf{(Comparison of Gold vs. retrieved evidence)}, all models, including baselines perform better with gold evidence for both FEVER and SciFact datasets in terms of SARI and BARTScore (see Table \ref{tab:combined_results}). The use of gold evidence serves as an upper bound (oracle), as gold annotations are more relevant than retrieved evidence. While retrieved evidence generally underperforms gold evidence, the performance gap is often narrow. Our proposed M$_2$C variants give comparable results with both gold and retrieval evidences, which eliminates the need of gold annotated evidence for our method. Notably, for M$_2$C-DM and M$_2$C$^+$-DM, results with gold evidence are same.

In relation to \textbf{RQ-3 (Gold vs. retrieved evidence)}, all models perform better with gold evidence, which serves as an oracle setup due to its higher relevance. However, there is less difference in performance. Specifically, our proposed M$_2$C variants, M$_2$C$_\text{DM}$ and M$_2$C$^+_\text{DM}$, achieve comparable results with both gold and retrieved evidence, demonstrating the effectiveness of our method without relying on annotated evidence. Notably, for M$_2$C$_\text{DM}$ and M$_2$C$^+_\text{DM}$, the results are the same with gold evidence.

 % the effectiveness and robustness of our method without reliance on gold annotations
% 3. For both the dataset and both the llm, the retrieved evidence -based M2C and M2C+ performs better than when gold evidence been used. 

% \para{Qualitative Analysis}
To address \textbf{RQ-4} \textbf{(Correlation between retriever and performance)}, Figure \ref{fig:correlation} demonstrates a positive correlation between retrieval quality (nDCG@10) and downstream correction performance (SARI) for each retriever, indicating that higher-quality retrievals lead to more accurate corrections. As shown in Table~\ref{tab:retrieval_analysis}, retriever choice substantially affects correction quality--retrievers like MonoT5 and ColBERT retrieve the most relevant evidence for given claims, while Contriever achieves the highest overall nDCG@10. Therefore, retrievers vary in effectiveness depending on the topic or the claim type. Thus, motivated by these observations, we proposed M$_2$C$^+$,~an~ensemble-based strategy that aggregates evidence from multiple retrievers to harness their complementary strengths and reduce noise of individual retrievers.

\begin{table}[t]
\centering
\small
\begin{adjustbox}{width=0.9\linewidth}
\begin{tabularx}{\linewidth}{@{}Xrr@{}}
\toprule
\multicolumn{3}{c}{\textbf{Input Claim:} \bbox{One Dance} was by a \rbox{Mexican}.} \\
\multicolumn{3}{c}{\textbf{Ground Truth:} \bbox{One Dance} was by a \bbox{Canadian}.} \\
\midrule
\textbf{Retrieved evidences} & \textbf{nDCG} & Time\\
\midrule
\textbf{BM25:} The Jarabe Tapatío, better known internationally as ``The \rbox{Mexican Hat Dance}" ...  & 0.5767 & 20--25\\
\midrule
\textbf{Contriever:} Aspen Santa Fe Ballet ASFB is an \rbox{American} \rbox{contemporary dance} company... & \textbf{0.7465} & 270\\
\midrule
\textbf{RM3:} \bbox{One Dance} is a 2003 \bbox{Canadian} romantic \rbox{drama film} about ... & 0.5892 & 20--30\\
\midrule
\textbf{ColBERT:} \bbox{``One Dance"} is a song by \bbox{Canadian} rapper Drake from his fourth studio album... & 0.6590 & 192\\
\midrule
\textbf{MonoT5:} \bbox{``One Dance''} is a song by Drake, a \bbox{Canadian} rapper, singer, and songwriter... & 0.6122 & 250\\
\bottomrule
\end{tabularx}
\end{adjustbox}
\caption{\small Examples of top-most similar sentences retrieved using different rankers from the FEVER dataset. These evidence are used in prompts for correcting the input claims. Also, we report nDCG@10. The \bbox{blue} highlighted parts denote correctly retrieved phrases and \rbox{red} parts denote the incorrect phrases responsible for the erroneous prospective corrections. Reported retrieval times (in seconds) correspond to 100 claims.}
\label{tab:retrieval_analysis}
\end{table}
\begin{table}[t]
\centering
\small
\begin{adjustbox}{width=.95\columnwidth}
\begin{tabular}{@{}ll cc ccc@{}}
\toprule
% \rowcolor{yellow}
  &  & \multicolumn{5}{c}{Time}  \\ 
\cmidrule(l){3-7}
   & Memory &\multicolumn{2}{c}{Ranking} & \multicolumn{2}{c}{Model} & Total\\ 
\cmidrule(l){3-4}\cmidrule(l){5-6}   
  Model &  (max) & Tr & Infer & Tr & Infer &  \\ 
\toprule 
Z{\tiny ERO}F{\tiny EC}-DA & 8GB & - &2.70 & - &6.5 & 6.5 \\ 
T5-Distant  & 48GB &18.0 &2.70 & 2.10 &1.00 & 23.80 \\ 
C{\tiny OMP}E{\tiny DIT}  & 5GB & - &2.70 & 18.50 & 1.00 & 23.20 \\ 
M$_2$C$_{\text{DM}}$ & 48GB  & - &2.70 & - & 7.5 & 10.20 \\ 
M$_2$C$_{\text{DM}}^+$ & 48GB  & - & 7.52 & - & 30.00 & 37.52 \\ 
\bottomrule
\end{tabular}%
\end{adjustbox}
\caption{\small GPU memory and total time (in hours) for processing all claims on FEVER dataset with Qwen as base model.} 
\vspace{-1.5em}
\label{tab:func_time}
\end{table}

\begin{table}[t]
\centering
\small
\begin{adjustbox}{width=.95\columnwidth}
\begin{tabular}{@{}llcccccc@{}}
\toprule
Method & \rotatebox[origin=c]{70}{BM25} & \rotatebox[origin=c]{70}{RM3} & \rotatebox[origin=c]{70}{Contriever} & \rotatebox[origin=c]{70}{ColBERT} & \rotatebox[origin=c]{70}{MonoT5} & SARI & BART\\ 
\toprule
\multirow{9}{*}{\rotatebox[origin=c]{0}{M$_2$C$^+$}} & \cmark & \xmark & \xmark & \cmark & \cmark & 50.4832& -2.5809 \\
& \cmark & \xmark & \cmark & \xmark & \cmark & 50.4231 & -2.5774 \\
& \cmark & \xmark & \cmark & \cmark  & \xmark & 49.7552& -2.6009 \\
\cmidrule(r){2-8}
 &\cmark & \cmark & \cmark &  \cmark& \xmark & 49.7772& -2.5925 \\
 &\cmark & \cmark & \cmark &  \xmark& \cmark & 50.0614 & -2.5852 \\
 &\cmark & \cmark & \xmark &  \cmark& \cmark & 50.0763 & -2.5907 \\
 &\cmark & \xmark & \cmark &  \cmark& \cmark & \textbf{50.7141}& -2.5724 \\
 &\xmark & \cmark & \cmark &  \cmark& \cmark & 50.6295 & \textbf{-2.5701} \\
 \cmidrule(r){2-8}
& \cmark & \cmark  & \cmark &  \cmark & \cmark & 50.4485 & -2.5779 \\
\bottomrule
\end{tabular}%
\end{adjustbox}
\caption{\small Retriever sensitivity analysis for M$_2$C$^+$ on the FEVER dataset using Qwen as the base model.}
\label{tab:qwen_ret_sen}
\end{table}
\begin{table*}[t]
\centering
\small
\begin{adjustbox}{width=0.9\textwidth}
\begin{tabularx}{\textwidth}{@{}X X@{}}
\toprule
\multicolumn{2}{c}{\textbf{Example 1 (\textsc{Majority Voting Failure})}} \\
\midrule
\textit{Input:} ``There are currently 417 Mormon members as of 2017.'' &
\textit{Ground Truth:} ``There are currently 15,882,417 Mormon members as of 2017.'' \\
One of the Candidate Answers: ``There are currently 15,882,417 Mormon members as of 2017.'' & 
\textit{Final Answer:} ``There are 64,123 members as of 2017.'' \\
\multicolumn{2}{c}{\textit{Note:} The correct candidate is present but outvoted.}\\
\midrule
\multicolumn{2}{c}{\textbf{Example 2 (\textsc{Entity Mismatch in Correction})}}\\
\midrule
\textit{Input:} ``LinkedIn is based in Spain.'' &
\textit{Ground Truth:} ``LinkedIn is based in the United States.''\\
& \textit{Final Answer:} ``LinkedIn is not based in Spain.''  \\
\multicolumn{2}{c}{\textit{Note:} Negation is correct, but it fails to provide the correct entity.}\\
\midrule
\multicolumn{2}{c}{\textbf{Example 3 (\textsc{Correction Scoring Failure})}}\\
\midrule
\textit{Input:} ``The Giver is a TV show.'' &
\textit{Ground Truth:} ``The Giver is a film.''  \\ 
Candidate Correction 1: ``The Giver is a TV show.'' & DocNLI: 0.975, ROUGE-L: 1.0 \\  
Candidate Correction 2: ``The Giver is a film.'' & DocNLI: 0.997, ROUGE-L: 0.767 \\
& \textit{Final Answer:} ``The Giver is a TV show.'' \\
\multicolumn{2}{c}{\textit{Note:} Correct option has higher factual score but loses due to lower ROUGE.}\\
\bottomrule
\end{tabularx}
\end{adjustbox}
\caption{Representative error cases in M$_2$C$^+$: failures due to majority voting, entity mismatch, and limitations of the correction scoring scheme.}
\label{tab:error_analysis}
\end{table*}

\subsection{Sensitivity Analysis} \label{sec:sensitivity_anal}
\para{Sensitivity of the Number of Examples}
% \payel{plot graph, (x->number of examples vs y->SARI score) for model (M$^2$C-KP(5 retrievers), M$^2$C$^+$-KP, M$^2$C$\text{W/OE}$-KP, 0-shot, RAG) --> scifact, fever --> qwen - 2 plots- 1 for fever, another for scifact}
In FEVER dataset, we observe from Figure~\ref{fig:fever_sensitivity},  M$_2$C$^+$ exhibits greater stability with respect to context size $p$ (see Equation \ref{eq:rag}) (i.e., the number of retrieved examples), consistently outperforming all individual retriever variants of M$_2$C and simple 0-shot. In contrast, for the SciFact dataset, M$_2$C$^+$ underperforms compared to M$_2$C with RM3 and MonoT5 as context size increases, likely due to domain-specific vocabulary and token mismatches affecting retrieval relevance (see Figure \ref{fig:sci_sensitivity}). Also for SciFact, across most experiments, the optimal performance is observed for $p=3$, which we have reported in Table~\ref{tab:combined_results}. However, for both datasets, M$_2$C with gold annotated evidence, i.e., M$_2$C-Gold, being an oracle setup, achieves the highest performance.

% 1. Figure~\ref{fig:fever_sensitivity}, In SciFact, M$_2$C$^+$ downperforms M$_2$C with RM3 and MonoT5 as the number of examples increases, possible reason due to domain-specific vocabulary and token mismatch. 

\para{Case Study on Scoring Function}
Figure \ref{fig:score_func} represents an ablation of the impact of different scoring configurations on the FEVER test set with a balancing factor of $\lambda = 0.5$ (see Appendix \ref{appx:lambda}). Inspired by \citet{manakul2023selfcheckgpt}, we use DeBERTa-v3-large~\cite{he2021debertav3} fine-tuned to MNLI as an entailment model with the ROUGE-L metric. Following the metric combinations used by \citet{huang2023zero}, we have compared the performance of the correction scoring in Figure \ref{fig:score_func}. We observe that combining DocNLI~\cite{yin-etal-2021-docnli} and ROUGE-L performs best in terms of SARI and BART score.

% with a balancing factor of λ = 0.5. We observe that the DocNLI + ROUGE-L combination performs best, achieving the highest SARI score. 
% \cite{manakul2023selfcheckgpt} we use DeBERTa-v3-large
% (He et al., 2023) fine-tuned to MNLI as the NLI
% model.

\para{Computational Analysis}
% Table~\ref{tab:qual_analysis}(b) reports the end-to-end inference time and GPU memory consumption for the proposed framework. Since the retrieval component is CPU-based and the indexing is performed once per corpus, minimizing recurring computational overhead.
% As an inference-only framework, M$_2$C and its ensemble variant M$_2$C$^+$ incur lower overall cost than existing training-based baselines (See Table \ref{tab:func_time}). Although M$_2$C$^+$ introduces additional latency due to multi-retriever aggregation, both variants remain efficient and practical for research-scale and moderately resource-limited settings. Hence, our framework avoids significant memory or latency bottlenecks and is well-suited for real-time scenarios. \textcolor{red}{Comparable results}

% The proposed M$_2$C framework demonstrates efficient computational behavior. While M$2$C${\text{DM}}^+$ incurs slightly higher latency due to multi-retriever aggregation, it provides only marginal gains over M$2$C${\text{DM}}$, indicating that the faster variant remains highly effective. With CPU-based retrieval and one-time indexing, both models are inference-only, require practical GPU memory, and avoid the heavy training overhead of baselines.
%
Table~\ref{tab:func_time} shows GPU memory and total processing time for all claims on FEVER. We observe that although, M$_2$C${_\text{DM}}^+$ incurs a moderate latency increase (2.7–4$\times$) than M$_2$C${_\text{DM}}$ due to multi-retriever aggregation, it provides only marginal gains over M$_2$C${_\text{DM}}$, indicating that the faster variant remains highly effective. The retrieval component is CPU-based, and the indexing is performed once per corpus, minimizing recurring computational overhead. Both are inference-only and maintain GPU usage (48GB), avoiding the heavy training overhead like baselines. Overall, our framework provided an efficient balance between computational cost and correction effectiveness.

% Table \ref{tab:func_time}
% -- points: combedit - inference (checkpoint exists)\\
% t5- training + inference (report)\\
% Table \ref{tab:qual_analysis}(b) -- retrieval time\\
\para{Sensitivity of Masking Strategies}
We analyze key hyperparameters for both masking strategies: the token masking ratio in RM, and the relevance–diversity trade-off ($\alpha$, Eq.~\ref{eq:mmr}) along with the number of selected entities ($m$) in DM. For RM, masking 15\% of tokens yields the best performance (Figure~\ref{fig:rm_sensitivity}), consistent with BERT~\cite{devlin2019bert}, while higher ratios degrade performance due to information loss. For DM, $\alpha=0.3$ provides the optimal trade-off between relevance and diversity (Figure~\ref{fig:DM_alpha}), while number of masked sentences in M$_2$C$_\text{DM}$, $m=10$ achieves the best balance between informativeness and noise reduction (Figure~\ref{fig:DM_sent}). We use these setting for all the subsequent experiments.

\para{Sensitivity of retrievers in M$_2$C$^+$} \label{sec:sensitivity_ret}
Table~\ref{tab:qwen_ret_sen} reports the sensitivity of M$_2$C$^+$ to different retriever combinations on the FEVER using Qwen as the base model. The results show that dense retrievers such as Contriever, ColBERT, and MonoT5 consistently strengthen the ensemble by retrieving semantically similar evidence, particularly under low lexical overlap. But, the inclusion of RM3 often underperforms due to its reliance on pseudo-relevance feedback. Also, combinations excluding RM3 but incorporating at least two dense retrievers provide the best results. The best SARI score is achieved when combining all retrievers except RM3, while excluding BM25 yields the best BART score. All the subsequent experiments are reported without RM3. As M$_2$C$^+$ is a majority voting based method, at least three retrievers are required to reach a decision; hence, we progressively constructed an ensemble starting with any three retrievers, then four, and finally all five. Thus, ensembling diverse retrievers effectively mitigates individual retriever biases and enhances downstream performance.

% Random masking ratio- Figure \ref{fig:rm_sensitivity} - best for 15\% , among 0.15,0.3,0.4,0.5 - we picked 0.15 for our experiments of random baseline masking.\\
% Diversity aware masking - \\
% vary $\alpha$ Table\ref{tab:alpha_da} - best is coming with $\alpha=0.3$, alpha balances relevance and diversity in MMR algorithm (explained in Section \ref{sec:masker}), we tried 0.1,0.3,0.5,0.7,0.9. Likely reason??.  \\
% vary sentences  table\ref{tab:no_sent_da}, this is m from the MMR selected newly reranked entites. So if we change the number of entites to consider in case of each input claim, we see for m=10, the result is best for fever data, on our model m2c. so we carried in all our experiments.

\subsection{Error Analysis}
Table \ref{tab:error_analysis} shows some examples where our model failed. Example~1 shows the limitation of the~majority voting approach, i.e., M$_2$C$^+$. Here, the desired correction was generated by a few retrievers but was ultimately discarded due to being outvoted by the erroneous prospective corrections from others. In example 2, the model correctly identifies that the claim is wrong (e.g., LinkedIn is not based in Spain) but fails to replace the incorrect part with the correct information (i.e., `the United States'). Although our model produced a logically valid correction, the model deviated from the desired output. Example 3 shows the failure of the correction scoring function (see Section \ref{sec:corr_score}). The model chooses a less factual candidate because it has a higher ROUGE score, even though the correct candidate has more factual alignment. By addressing the genesis of these errors we can make the factual correction model more reliable.

Some errors arise from annotation inconsistencies in the benchmark datasets, causing the evaluation to unfairly penalize our model’s outputs; we discuss these cases in detail in Appendix~\ref{appx:annotation}.
% So, although our model produced a logically valid correction, the reliance on negation instead of entity replacement led to a drop in performance.

% Our error analysis, based on Table~\ref{tab:error_analysis}, identifies three main types of failures in our factual correction system. First, in majority voting failure, the correct correction is present among the candidates but gets outvoted, leading to an incorrect final prediction—for example, the true number of Mormon members is overlooked. Second, in entity mismatch, the model correctly identifies that the claim is wrong (e.g., LinkedIn is not based in Spain) but fails to replace it with the correct information (i.e., the United States).
% Third, in correction scoring failure, the model chooses a less factual candidate because it has a higher ROUGE score, even though the correct candidate has a stronger factual alignment based on DocNLI. These errors show the limitations of majority voting, lack of entity grounding, and over-reliance on surface-level similarity metrics like ROUGE. Improving voting strategies, entity replacement, and fact-aware scoring can help make factual correction more reliable.

\section{Conclusions and Future Work}

% In this paper, we propose a training-free, annotation-free, RAG-based framework Mask-to-Correct (M$_2$C), for faithful fact correction. Unlike traditional masking methods, in M$_2$C we introduced a diversity-aware masking strategy for selecting diversified but relevant entities. Furthermore, we propose an ensemble approach, M$_2$C$^+$, aggregating correction candidates across diverse retrievers. Extensive experiments on FEVER and SciFact datasets demonstrate that our method consistently outperforms several parametric and non-parametric baselines. In future, to improve transparency and trustworthiness, we plan to extend M$_2$C to a self-correcting framework by incorporating a feedback loop, where generated corrections will guide retriever reranking and masking.

We propose Mask-to-Correct (M$_2$C), a training and annotation-free RAG-based framework for faithful fact correction. Here, we introduce a diversity-aware masking strategy for selecting diversified but relevant entities. Furthermore, we propose an ensemble approach, M$_2$C$^+$, that aggregates corrections across diverse retrievers. Extensive experiments on FEVER and SciFact datasets demonstrate that our method consistently outperforms all baselines. In future, to improve transparency and trustworthiness, we plan to extend M$_2$C to a self-correcting framework by incorporating a feedback loop, where generated corrections will guide retriever reranking and masking.
% Thereby improving transparency and trustworthiness in real-world applications. 
% Additionally  low-resource languages 

% , we presented Mask-to-Correct (M2C), a training-free, retrieval-augmented framework for factual error correction that localizes erroneous spans in claims and corrects them using evidence-guided generation. To enhance robustness against retriever-specific biases, we further proposed M2C+, an ensemble-based extension that aggregates candidate corrections across diverse retrievers using majority voting.

% We introduced a novel diversity-aware masking strategy that selects correction spans based on both semantic relevance and diversity, which consistently outperforms traditional masking techniques across datasets. 

% Extensive experiments on FEVER and SciFact demonstrate that our approach not only outperforms both parametric and non-parametric baselines, but also maintains strong performance even without access to gold-standard evidence.
\vspace{-1mm}
% \section*{Limitations}

\para{Acknowledgment} We used Gen AI tools for minor paraphrasing. Final content was reviewed.

\section*{Limitation} 
A limitation is that the computation is memory-intensive as one needs to load the LLMs, thus making it difficult to execute our approach on a terminal with limited memory capacity. %constrained resource environment to utilize its full capacity.
Additionally, due to the limitations of our own resources, we were not successful in employing a larger or a commercially accessible LLM (e.g., GPT-4, Claude).

% \payel{write limitations}
% \section*{Ethical Considerations} 

% \bibliographystyle{cas-model2-names}
\bibliography{custom}

% \newpage
\appendix

\section{Metric Descriptions} \label{appx:metrics}
\subsection{SARI Score} SARI~\cite{xuoptimizing} is designed for tasks where a system must edit an input sentence (e.g., factual correction), rather than generate text from scratch. Unlike overlap-based metrics such as BLEU or ROUGE, SARI explicitly evaluates the goodness of the edits by comparing the input sentence, system output (predicted correction), and reference (i.e., ground truth or retrieved supporting context).

SARI is computed using three components as, 
\begin{equation*}
    \text{SARI}=\frac{1}{3}(\text{F}_\text{add}+\text{F}_\text{keep}+\text{P}_\text{del})
\end{equation*}
where, $\text{F}\text{add}$ assesses the addition of relevant n-grams present in the reference but absent in the input, $\text{F}\text{keep}$ evaluates the preservation of correct input n-grams, and $\text{P}_\text{del}$ measures the removal of incorrect or unnecessary content. In our experiments, we use $n=4$ for n-gram computation. 

By utilizing add, keep, and delete operations, SARI rewards only edits that align with ground-truth corrections and penalizes incorrect or excessive rewriting, making it well-suited for factual correction tasks where minimal yet accurate changes are desired.

% \begin{itemize}
%     \item $\text{F}_\text{add}$ measures whether the system adds n-grams that appear in the reference but not in the input, rewarding insertion of missing correct facts and penalizing unsupported additions. 
%     \item $\text{F}_\text{keep}$ measures whether the system preserves the correct n-grams from the input, rewarding retention of true information and penalizing unnecessary changes.
%     \item $\text{P}_\text{del}$ measures whether the system correctly deletes n-grams that should not remain in the corrected output, rewarding removal of incorrect content.
% \end{itemize}
% For our experiment, we took $n=4$ for n-gram.

\subsection{BARTScore}
BARTScore~\cite{yuan2021bartscore} is a reference-based evaluation metric that measures the relationship between two text sequences using a pretrained sequence-to-sequence model. It computes the log-likelihood of generating a target sequence conditioned on a source sequence, thereby capturing how well one text is supported by another.

Formally, given an input claim $\vec{x}$ and a reference evidence sentence $e$, the BARTScore is defined as:
\begin{equation*}
    \text{BARTScore}(\vec{x}, e) = \frac{1}{|e|} \sum_{t=1}^{|e|} \log P(e_t \mid e_{<t}, \vec{x}),
\end{equation*}
where $P(\cdot)$ is computed using a pretrained BART model.

In our work, we use BARTScore to measure how well our correction is supported by the gold evidence. Specifically, the score reflects the extent to which the corrected claim can account for or generate the supporting evidence, thereby indicating its factual grounding.

\begin{table}[h]

\small
\begin{adjustbox}{width=0.9\columnwidth}
\begin{tabularx}{.98\columnwidth}{@{}X@{}}
\toprule
\gc{\textbf{Annotation Error}} \\
\midrule
\textbf{Example 1}\\
\midrule
\textbf{Input Claim:} Exit the King is by  \rbox{a man}.\\
\textbf{Ground Truth:} Exit the King is by  \rbox{a man}. \\
% \midrule
\noindent\dotfill
\newline
\textbf{Our Correction:} Exit the King is by \bbox{Eugène Ionesco}. \\
\textbf{Supporting Evidence:} \bbox{Exit the King} is a play by \bbox{Eugène Ionesco}, originally written in French in 1962.
\\
\midrule
\textbf{Example 2}\\
\midrule
\textbf{Input Claim:} A U.S. state is \rbox{Phoenix, Arizona}.\\
\textbf{Ground Truth:} \rbox{Phoenix, Arizona} is a U.S. \rbox{state}. \\
% \midrule
\noindent\dotfill
\newline
\textbf{Our Correction:} A U.S. \bbox{state is Arizona}. \\
\textbf{Supporting Evidence:} \bbox{Arizona is a state} located in the southwestern region of the United States.\\
\bottomrule
\end{tabularx}
\end{adjustbox}
\caption{\small
Example of an annotation error in the dataset. The \bbox{blue}-highlighted spans indicate correct factual information, while the \rbox{red} highlights indicate incorrectly accepted text in the original claim and annotation.}
\label{tab:annotation_error}
\end{table}

\begin{table*}[t]
\centering
\small
\begin{adjustbox}{width=\textwidth}
\begin{tabular}{p{3.5cm} p{3cm} p{3.5cm} p{5cm} c}
\toprule
\textbf{Input Claim} & \textbf{GT Claim (TRUE)} & \textbf{Model Output} & \textbf{Observation} & \textbf{SARI} \\
\midrule

Jenny McCarthy modeled for Playboy photographers. & 
Same as input & 
Jenny McCarthy modeled for Playboy magazine. & 
Semantics preserved; minor paraphrase penalized. & 
28.79 \\

\midrule

An American directed One True Thing. & 
Same as input & 
Carl Franklin directed One True Thing. & 
Output is more specific and factually correct; SARI penalizes ADD. & 
22.64 \\

\midrule

The left hand side of the Trebbia River… & 
Same as input & 
The left bank of the Trebbia River… & 
Bank vs side; paraphrased → SARI penalizes. & 
28.82 \\

\midrule

Great white sharks have never killed people. & 
Same as input & 
Great white sharks have killed people. & 
GT claim is factually incorrect, so SARI unfairly penalizes correction. & 
22.45 \\

\midrule

Exit the King is by man. & 
Same as input & 
Exit the King is by Eugène Ionesco. & 
GT claim is incorrect and this claim is ambiguous. & 
28.29 \\

\bottomrule
\end{tabular}
\end{adjustbox}
\caption{\small Examples demonstrating the limitations of SARI on TRUE-labelled instances (input equals ground truth). The metric penalizes factually correct outputs due to its sensitivity to surface-form changes, including valid paraphrases, informative refinements, and inconsistencies or errors in the ground-truth annotations.}
\label{tab:sari_limitations}
\end{table*}

\begin{table}[h]
\centering
\small
\begin{adjustbox}{width=.99\columnwidth}
\begin{tabular}{@{}llcccrr@{}}
\toprule
 & \multicolumn{2}{c}{SARI (\%)} & \multicolumn{2}{c}{BART}  & \multicolumn{2}{c}{Overall} \\ 

 \cmidrule(l){2-3}  \cmidrule(l){4-5} \cmidrule(l){6-7}
Model & TRUE & FALSE & TRUE & FALSE  &  SARI & BART\\ 
\toprule				
M$_2$C& 31.5323  & 63.4833 & -2.6904 & \textbf{-2.4995}  & 50.3638 &	-2.6848\\ 
M$_2$C$^+$    & \textbf{34.1467} & \textbf{64.0014} & \textbf{-2.6905}  & -2.4900& \textbf{50.7141}& \textbf{-2.5724} \\ 
\bottomrule
\end{tabular}%
\end{adjustbox}
\caption{\small Class-wise performance of M$_2$C (MonoT5 as retriever) and M$_2$C$^+$ on the FEVER dataset with Qwen as base LLM. Bold values indicate the best performance between M$_2$C and M$_2$C$^+$.} 
% The top part corresponds to large datasets, and the bottom to small datasets.
\label{tbl:classwise_result}
\end{table}

\subsection{DocNLI} DocNLI~\cite{yin-etal-2021-docnli} is a document-level natural language inference task in which a model determines whether a claim $\vec{x}$ is entailed by, contradicted by, or unsupported with respect to an evidence document $e \in \mathcal{E}(\vec{x})$. Formally, given a pair $(\vec{x},e)$, the classifier outputs one of three labels: $y \in \{\text{entailment},\text{contradiction},\text{neutral}\}$.

In our work, we use a RoBERTa~\footnote{\url{https://huggingface.co/FacebookAI/roberta-large}} model fine-tuned on the DocNLI dataset and use the entailment probability, i.e., 
\begin{equation*}
    \text{Entail}(\vec{x},e)=P(y= \text{entailment}|\vec{x},e),
\end{equation*}
as a factuality score for candidate corrections. In this paper, this score captures the degree of factual alignment between the corrected claim and the retrieved evidence, and is used as a component of the correction scoring function to rank candidate outputs based on their entailment likelihood (as described in Section \ref{sec:corr_score}).

\subsection{ROUGE-L}
ROUGE-L~\cite{lin2004rouge} measures the similarity between a generated sentence and a reference based on the length of their Longest Common Subsequence (LCS), capturing both lexical overlap and word order.

Formally, given a candidate correction $\vec{x}'$ and the original claim $\vec{x}$, the ROUGE-L score is computed using LCS-based precision and recall, which are combined into an F1-score.

In our work, we use ROUGE-L as a proxy for edit minimality, i.e., to quantify how much the corrected candidate deviates from the input claim. Similar to \textit{ZeroFEC}~\cite{huang2023zero}, we use this score in conjunction with the DocNLI-based entailment score to rank candidate corrections.

Specifically, while the DocNLI score captures factual consistency with respect to external evidence, ROUGE-L encourages minimal and conservative edits by favoring candidates that preserve the structure and content of the original claim. The combination of these two signals ensures that the selected correction is both factually grounded and minimally deviating from the input, thereby maintaining faithfulness while correcting errors.
% \section{Experiment Environment}: --As correct paper

\section{Dataset Description} \label{appx:data_desc}
In this paper, we conduct experiments on two benchmark datasets: FEVER~\cite{thorne-vlachos-2021-evidence} and SciFact~\cite{wadden-etal-2020-fact} datasets. We use these fact correction datasets introduced by \citep{huang2023zero}, where the supported claims are taken as faithful, and unfaithful versions are generated by applying Knowledge Base Informed Negations~\cite{wright-etal-2022-generating} to a subset of the faithful claims. Moreover, we used the FEVER validation data~\cite{thorneEvi} to conduct an \textbf{extensive grid search} to optimize the hyperparameters for our downstream fact correction task. 
% \input{table/table_data}

% \begin{figure}[t]
% \centering
% \begin{adjustbox}{width=0.93\columnwidth}
% {\small
% \begin{tcolorbox}[colback=gray!10, colframe=black, title=Prompt Template, arc=2pt]
% \texttt{Given a claim and related evidence, your task is to correct it if the claim is not supported by the given evidence. If the input claim is correct, do not edit it and give the input claim as output. Your output claim should be faithful to the evidence and should not deviate much from the input claim. Do not print anything else in the output except the corrected claim. Strictly follow the syntax given below for output syntax:
% \newline Input Claim: [sentence]
% \newline Evidence: [document]
% \newline Output Correction: [sentence] }
% \
% \end{tcolorbox}
% }
% \end{adjustbox}
% \begin{adjustbox}{width=0.93\columnwidth}
% {\small
% \begin{tcolorbox}[colback=gray!10, colframe=black, title=Selected Examples, arc=2pt]
% \texttt{[1.] Pulmonary embolism (PE) is a blockage of an artery in the lungs by...Signs of a PE include low blood oxygen levels.\newline
% [2.]...}
% \end{tcolorbox}
% }
% \end{adjustbox}
% \begin{adjustbox}{width=0.93\columnwidth}
% {\small
% \begin{tcolorbox}[colback=gray!10, colframe=black, title=Test Instance, arc=2pt]
% \texttt{ Input Claim: Pulmonary embolism is indicated by high blood oxygen levels.\newline
% Output Correction:}
% \end{tcolorbox}
% }
% \end{adjustbox}
% \caption{An illustration of the prompt structure used in the RAG (few-shot) experiment.}
% \label{fig:prompt_template}
% % \input{fig_def/prompt_temp}
% \end{figure}

\begin{figure*}[t]
\centering
\begin{adjustbox}{width=0.87\textwidth}
{\small
\begin{tcolorbox}[colback=gray!10, colframe=black, title=Prompt Template, arc=2pt]
\texttt{Given a claim and related evidence, your task is to correct it if the claim is not supported by the given evidence. If the input claim is correct, do not edit it and give the input claim as output. Your output claim should be faithful to the evidence and should not deviate much from the input claim. Do not print anything else in the output except the corrected claim. Strictly follow the syntax given below for output syntax:
\newline Input Claim: [sentence]
\newline Evidence: [document]
\newline Output Correction: [sentence]}
\end{tcolorbox}
}
\end{adjustbox}

\vspace{0.5em}

\begin{adjustbox}{width=0.87\textwidth}
{\small
\begin{tcolorbox}[colback=gray!10, colframe=black, title=Selected Examples, arc=2pt]
\texttt{[1.] Pulmonary embolism (PE) is a blockage of an artery in the lungs by... Signs of a PE include low blood oxygen levels.\newline
[2.] ...}
\end{tcolorbox}
}
\end{adjustbox}

\vspace{0.5em}

\begin{adjustbox}{width=0.87\textwidth}
{\small
\begin{tcolorbox}[colback=gray!10, colframe=black, title=Test Instance, arc=2pt]
\texttt{Input Claim: Pulmonary embolism is indicated by high blood oxygen levels.\newline
Output Correction:}
\end{tcolorbox}
}
\end{adjustbox}

\caption{An illustration of the prompt structure used in the few-shot RAG experiment.}
\vspace{-1em}
\label{fig:prompt_template}
\end{figure*}

\section{Information of our Annotated data} \label{appx:annotation}

Table \ref{tab:annotation_error} illustrates examples of annotation errors in the benchmark dataset. In Example 1, the input claim ``Exit the King is by a man'' is marked as correct, although the accurate correction should be Eugène Ionesco; here, “man” is a generalized term. Similarly, in Example 2, the claim ``A U.S. state is Phoenix, Arizona'' is truth claimed by annotators, but originally Phoenix is a city in the state of Arizona. In both case, our model correctly generates predicted outputs using retrieved evidence, but is penalized due to wrong gold annotations.
% Table \ref{tab:annotation_error}
% The input claim was incorrectly annotated as correct, though the actual answer refers to the playwright Eugène Ionesco.

% These errors underscore the limitations of manual annotation and reveal the impact of pooling bias, which can misrepresent true model performance in fact correction tasks.

% In both examples, the ground-truth (GT) annotations incorrectly accept factually inconsistent claims. In Example 1, the input claim “Exit the King is by a man” is marked as correct, though the accurate correction should reference the actual playwright Eugène Ionesco. Similarly, in Example 2, the claim “A U.S. state is Phoenix, Arizona” is accepted, despite being a reversed form of the factual statement that Phoenix is a city in the state of Arizona. While our model correctly generates factually consistent outputs using retrieved evidence, it is penalized due to erroneous gold annotations. These cases highlight the impact of pooling bias in crowd-annotated datasets, where insufficient evidence exposure or annotation oversight can mislabel incorrect claims as correct—ultimately affecting the fair evaluation of fact correction models.

\section{Real-World Applicability} \label{appx:real_WORLD}
FEVER and SciFact include both true and false as well as ambiguous claims resembling real-world scenarios. For instance, examples such as ``Parkinson's disease causes symptoms", ``Shut Up is a title", ``Benzodiazepines can be taken", and ``Fraud can be used for gain" are ambiguous, context-dependent claims and require the retrieval of appropriate evidence for correction. In practical settings, a fact-correction system must not only revise false claims but also preserve correct ones and appropriately handle ambiguous cases. To evaluate this behavior, we analyze the effectiveness of our framework across different classes, as shown in Table~\ref{tbl:classwise_result}.

We observe that, M$_2$C and M$_2$C$^+$ effectively preserve correctness for true claims while correcting false ones, thus the framework does not operate under the assumption that all inputs are incorrect and hence generalizing well in real-world scenarios. 

Moreover, we observe that $\text{SARI} (True) < \text{SARI} (False)$. This arises from the nature of SARI as an edit-based metric, which rewards necessary modifications but penalizes unnecessary changes. Since FALSE claims require edits, they achieve higher scores, whereas TRUE claims ideally remain unchanged, leading to penalization even for minor paraphrases.

As shown in Table~\ref{tab:sari_limitations}, our manual inspection reveals that the observed performance degradation is sometimes due to annotation inconsistencies and underspecified ground-truth claims, which cause SARI to penalize factually correct outputs.

% reveals that annotation inconsistencies and underspecified ground truth can cause SARI to penalize correct outputs, leading to an underestimation of model performance; examples are shown in Table~\ref{tab:sari_limitations}.

% To further understand this discrepancy, we manually inspect some instances and find that annotation inconsistencies and underspecified ground-truth claims often cause SARI to penalize factually correct outputs. These findings suggest that, although our model behaves appropriately in real-world settings by preserving correct claims, evaluation using SARI may underestimate its effectiveness. Detailed examples are provided in Table~\ref{tab:sari_limitations}.

\section{Implementation Details} \label{appx:exp_env}
We conduct all our experiments using the LLM Llama-2.0\footnote{\scriptsize \url{https://huggingface.co/TheBloke/Llama-2-70B-Chat-AWQ}} (70B)~\citep{touvron2023llama} and Qwen-2.5\footnote{\scriptsize \url{https://huggingface.co/Qwen/Qwen2.5-32B-Instruct-AWQ}} (32B)~\citep{qwen2} model. For our proposed diversity-aware masking-based methods, like M$_2$C$_\text{DM}$ and M$_2$C$^+_\text{DM}$, we used all-MiniLM-L6-v2 for getting embeddings of entities and used KeyBERT\footnote{\scriptsize \url{https://github.com/MaartenGr/KeyBERT}} \cite{grootendorst2020keybert,bennani2018simple}. For entity-masker, we extract the named entities using SpaCy\footnote{\url{https://spacy.io/}} \cite{vasiliev2020natural} and verb, adjective, and noun phrases using Stanza\footnote{\url{https://stanfordnlp.github.io/stanza/}} \cite{qi2020stanza}, and subsequently mask them. 
Next, for all our LLM-based M$_2$C and M$_2$C$^+$ experiments, we utilize the vLLM~\citep{kwon2023efficient,lin2024awq} library to apply k-v cache optimization, enhancing computation speed. For fine-tuning the supervised baselines in our experiments (namely, T5-distant, C{\tiny OMP}E{\tiny DIT}, and Z{\tiny ERO}F{\tiny EC}-DA), we follow the respective setups as reported in their original works. T5-distant uses a heuristic masking strategy and is fine-tuned on randomly masked data for 10 epochs with a learning rate of $5e-5$. C{\tiny OMP}E{\tiny DIT} employs the publicly available BART-large~\cite{lewis2019bart} post-editor trained for 10 epochs with batch size 64 on compression data generated using a BART-based perturber, and inference is performed using greedy decoding~\cite{germann2003greedy}. In our implementation, we used SpaCy~\cite{vasiliev2020natural} and Stanza~\cite{qi2020stanza} for entity recognition.  For Z{\tiny ERO}F{\tiny EC}-DA, we use the domain adaptation variant, where the DocNLI model is fine-tuned on P{\tiny UB}M{\tiny ED}QA~\cite{jin-etal-2019-pubmedqa} and B{\tiny IO}ASQ~\cite{tsatsaronis2015overview} datasets for up to 5,000 steps using AdamW~\cite{loshchilov2018decoupled} with a learning rate of $3e-6$. All generative models use beam search~\cite{freitag-al-onaizan-2017-beam} with a beam width of 4 during inference. All experiments throughout the paper were performed using a NVIDIA A6000 (48GB) GPU.

\begin{table}[h]
\centering
\small
\begin{adjustbox}{width=.98\columnwidth}
\begin{tabular}{@{}llcccc@{}}
\toprule
 & \multicolumn{5}{c}{$\lambda$}\\ 
\cmidrule(l){2-6}
% \cmidrule(l){4-5}
Metric & 0.0 & 0.2 & 0.5 & 0.8 & 1.0 \\ 
\toprule
SARI (\%) & 21.0695& 40.7768 &\textbf{52.7064}  & 51.7891 & 50.5760\\
%BARTScore & &  &  &  &\\
\bottomrule
\end{tabular}%
\end{adjustbox}
\caption{Ablation study on the effect of the balancing factor $\lambda$ in our M$_2$C framework using MonoT5 ranker on the FEVER validation set with Qwen model. The table reports SARI scores (\%) for different values of $\lambda$.}

% The top part corresponds to large datasets, and the bottom to small datasets.
\label{tbl:lambda}
\end{table}
\begin{figure*}[t]
\centering
% \begin{adjustbox}{width=0.93\columnwidth}
\begin{adjustbox}{width=0.87\textwidth}
{\small
\begin{tcolorbox}[colback=gray!10, colframe=black, title=Prompt Template, arc=2pt]
\texttt{Your task is to correct a claim by filling in the [MASK] using the provided input evidence, ensuring that the corrected claim is supported by the evidence and only differs from the input claim in the masked positions. If the input claim is correct, do not edit it and give the input claim as output. Your output claim should be faithful to the provided evidence and should not deviate much from the input claim.
\newline
Please use the most relevant evidence to correct the claim. The corrected claim shouldn't contain any [MASK].
\newline
\newline Input Claim: [sentence]
\newline Evidence: [document]
\newline Masked Claim: [masked sentence]
\newline Output Correction: [sentence] }
\
\end{tcolorbox}
}
\end{adjustbox}
% \begin{adjustbox}{width=0.93\columnwidth}
\begin{adjustbox}{width=0.87\textwidth}
{\small
\begin{tcolorbox}[colback=gray!10, colframe=black, title=Selected Examples, arc=2pt]
\texttt{[1.] Pulmonary embolism (PE) is a blockage of an artery in the lungs by...Signs of a PE include low blood oxygen levels.\newline
[2.] ...}
\end{tcolorbox}
}
\end{adjustbox}
% \begin{adjustbox}{width=0.93\columnwidth}
\begin{adjustbox}{width=0.87\textwidth}
{\small
\begin{tcolorbox}[colback=gray!10, colframe=black, title=Test Instance, arc=2pt]
\texttt{Input Claim: Pulmonary embolism is indicated by high blood oxygen levels.\newline
Masked Claim: Pulmonary embolism is indicated by [MASK].\newline 
Output Correction: 
}
\end{tcolorbox}
}
\end{adjustbox}
\caption{An illustration of the prompt structure used in our proposed approach M$_2$C. }
\vspace{-1em}
\label{fig:m2c}
\end{figure*}

\section{Ablation on the balancing factor $\lambda$} \label{appx:lambda}
We study the impact of the balancing factor $\lambda$, which controls the trade-off between factuality and faithfulness. As shown in Table~\ref{tbl:lambda}, $\lambda = 0.5$ gives the best performance in terms of SARI score for M$_2$C-MonoT5 experiment. Thus we adopt this value for all experiments throughout the paper.

\section{Prompt Templates} \label{appx:prompt_temp}
Figure \ref{fig:prompt_template} shows the template used for the 0-shot and RAG experiment, and Figure \ref{fig:m2c} shows the templates we used in the evidence-guided corrector part of M$_2$C framework.

% Table~\ref{tab:qwen_ret_sen} presents the sensitivity analysis of M$_2$C$^+$ with different retriever combinations on the FEVER dataset using Qwen as the base model. Dense retrievers such as Contriever, ColBERT, and MonoT5 consistently improve ensemble performance by retrieving semantically aligned evidence, particularly under low lexical overlap. In contrast, RM3 tends to degrade performance due to noisy pseudo-relevance feedback. The best SARI score is achieved when combining all retrievers except RM3, while excluding BM25 yields the best BART score. Hence, all subsequent experiments are reported without RM3. Overall, the results demonstrate that ensembling diverse dense retrievers effectively mitigates individual retriever biases and enhances factual correction robustness.

\end{document}